\documentclass[article]{aa} 

\usepackage{natbib}
\usepackage{epsfig,latexsym,graphicx,epsf,subfigure}
\usepackage{threeparttable,hhline,longtable}
\usepackage{deluxetable} 
\usepackage{txfonts}

\newcommand{\eqref}[1]{(\ref{#1})}
\newcommand{\hsiao}{\textsc{hsiao}}
\newcommand{\salt}{\textsc{salt2}}

\def\lsim{\lesssim}
\begin{document}

\title{The colour-lightcurve shape relation of Type~Ia supernovae and
the reddening law}

\author{S.~Nobili\inst{1} \and A.~Goobar\inst{1}}
\authorrunning{Nobili \and Goobar}

\offprints{S.~Nobili, serena@physto.se}

\institute{Department of Physics, Stockholm University, S--106 91
  Stockholm, Sweden } 

\date{Received ...; accepted ...}

\date{} \abstract{} { A study of the time sequence of optical colours
of a large sample of nearby Type~Ia supernovae (SNe~Ia) is presented.
We study the dependence of the colour time evolution with respect to
the lightcurve shape, parametrized by the stretch factor.}  {We fit the spectral
template that minimizes the colour dispersion in SNe~Ia, as measured
through UBVRI photometry of near-by supernovae.}  {A clear colour
dependence upon lightcurve shape is found, with the narrower
lightcurves being redder up to about one month past lightcurve
maximum. We also derive an average reddening law, after correcting for
lightcurve shape differences in intrinsic colour, that is well
described by a Cardelli, Clayton \& Mathis law with $R_V=1.75 \pm
0.27$ for 80 Type Ia supernovae with $E(B-V) \le 0.7$ mag. A subset
sample including 69 SNe with modest reddening, $E(B-V)<0.25$ mag,
yields a significantly smaller value, $R_V \sim 1$, suggesting 
that the observed reddening of Type Ia supernovae may have a
more complex origin, perhaps involving other processes 
beside extinction by interstellar dust in the host galaxy.} {}
\keywords{supernovae: general - Stars: statistics}

\maketitle

\section{Introduction}
\label{sec:intro}
The use of type~Ia supernovae (SNe~Ia) as standardized candles has been
successfully exploited in recent years as a precise tool for cosmology
to discover the acceleration of the Universe through some yet undetermined
dark energy \citep{1998AJ....116.1009R,1999ApJ...517..565P}.
More recently we have seen the start of the second generation of
supernova surveys, with dedicated supernova searches (e.g
SNLS\footnote{http://www.cfht.hawaii.edu/SNLS/},
SDSS\footnote{http://www.sdss.org/},
ESSENCE\footnote{http://www.ctio.noao.edu/essence/}) and massive
spectroscopic follow up. The results from these surveys are showing
that increasing the number of supernovae populating the Hubble diagram
at high redshift may not be sufficient to discriminate among
alternative dark energy models. Systematic uncertainties appear to be
the limiting factors for the ongoing supernova efforts 
\citep{2006A&A...447...31A,2007astro.ph..1041W}.  

Dimming by dust along the line-of-sight, predominantly in the host
galaxy of the supernova explosion, is one the main sources of
systematic uncertainty, see e.g. \citet{2003ApJ...598..102K}.  
For Type Ia supernovae, two additional causes of reddening are
sometimes suggested: extinction by circumstellar dust and the
possibility of an intrinsic colour-brightness relation.   The last two
scenarios further complicate the prospects to disentangle the effects
and their possible evolution with cosmic time.  

Our ability to correct for these effects relies on the knowledge of
the intrinsic colours of SNe~Ia, as well as on the nature of the
reddening. The latter is often assumed to be similar to what is found
for extinction by dust in the Milky Way. As we will see in
section~\ref{sec:extlaw}, this is not necessarily the case, and needs
further investigation. 

The effect of cosmological redshift on the measured broadband filters
is accounted in a straight-forward manner through $K$-corrections
\citep{1996PASP..108..190K}.  However, the SN spectrum is usually not
measured at all epochs, since typically only few spectra are taken to
allow type identification. The standard technique consists in using
spectroscopic templates, built by averaging spectra of well
observed (mostly nearby) SNe~Ia. Thus, the uncertainty in
$K$-corrections depends primarily on the spectroscopic diversity of
SNe~Ia.  \citet{2002PASP..114..803N} showed that changes in individual
spectral features do no significantly affect $K$-corrections, 
as long as there is a good match between the observed and the rest
frame bandpass filters. These are instead mainly sensitive to the
supernova colours. Once again, the knowledge of intrinsic SN colour can
help to solve one of the limiting factors in supernova cosmology.

This issue becomes even more critical when dealing with the
observations of very high-z SNe, for which the rest frame extends,
partly or entirely, into the UV part of the spectrum. Our limited
knowledge of the supernova properties in the U-band, due to poor
telescope and CCD sensitivity at these wavelengths, was identified as
the main source of uncertainty in the determination of cosmological
parameters in \citet{2003ApJ...598..102K}.  

Some efforts to observe SNe in the U-band have been made and collected
a number of nearby SNe data \citep{2006AJ....131..527J,
2007MNRAS.tmp..161P,2007A&A...469..645S}.  At the same time there have
been attempts of using spectra of high redshift SN~Ia from the SNLS
sample, redshifted to optical wavelengths with good sensitivity, under
the assumption that there is no evolutionary trend in the SN spectrum
\citep{2005A&A...443..781G,2007A&A...466...11G,2007ApJ...663.1187H}.
This assumption may not hold up to arbitrary redshifts.  
As the average metallicity of the universe
increases with cosmic time, it is not unreasonable to expect that high
redshift SNe~Ia stem from environments with lower average
metallicity.  The effect on the spectral energy distribution of a
lower metallicity progenitor has been modeled by
\citet{1998ApJ...495..617H} and \citet{2000ApJ...530..966L} who
found that such SNe~Ia, especially at early epochs, are expected to
show enhanced flux in the UV region of the spectrum, weaker absorption
features in the optical and a shift in the minima of optical
absorption features to redder wavelengths (see, however, \citet{2007arXiv0710.3896E}).

In this paper we present a statistical study of $U-B$, $B-V$, $V-R$
and $R-I$ colour evolution for 80 nearby supernovae. This work is a
continuation of the analysis presented in \citet{2003A&A...404..901N},
with an increased sample and with the addition of $U$-band data. This
allowed us to study the dependence of colours on the stretch
factor. That is, in turn, applied to correct the spectral templates
used for computing the $K$-corrections.  The knowledge of intrinsic
colour dispersion is further used to study the average properties of the
extragalactic reddening law of SNe~Ia.

In section~\ref{sec:data} the data set used in this paper is
presented. In section~\ref{sec:average} we investigate the correlation
of colours with lightcurve shape, and estimate colour curve models for
typical SNe~Ia.  The study of intrinsic dispersion in colours is
reported section~\ref{sec:intrinsic} report. In
section~\ref{sec:template} we discuss the use of our colour-stretch
relation to modify the spectral template for computing $K$-correction,
and compare our results with previously published estimates.  Finally in
section~\ref{sec:extlaw}, we discuss the implication of our analysis
for computing the effective average extinction law of SNe~Ia.

\section{The data set}
\label{sec:data} 
\begin{figure}[htb]
\centering \includegraphics[width=8cm]{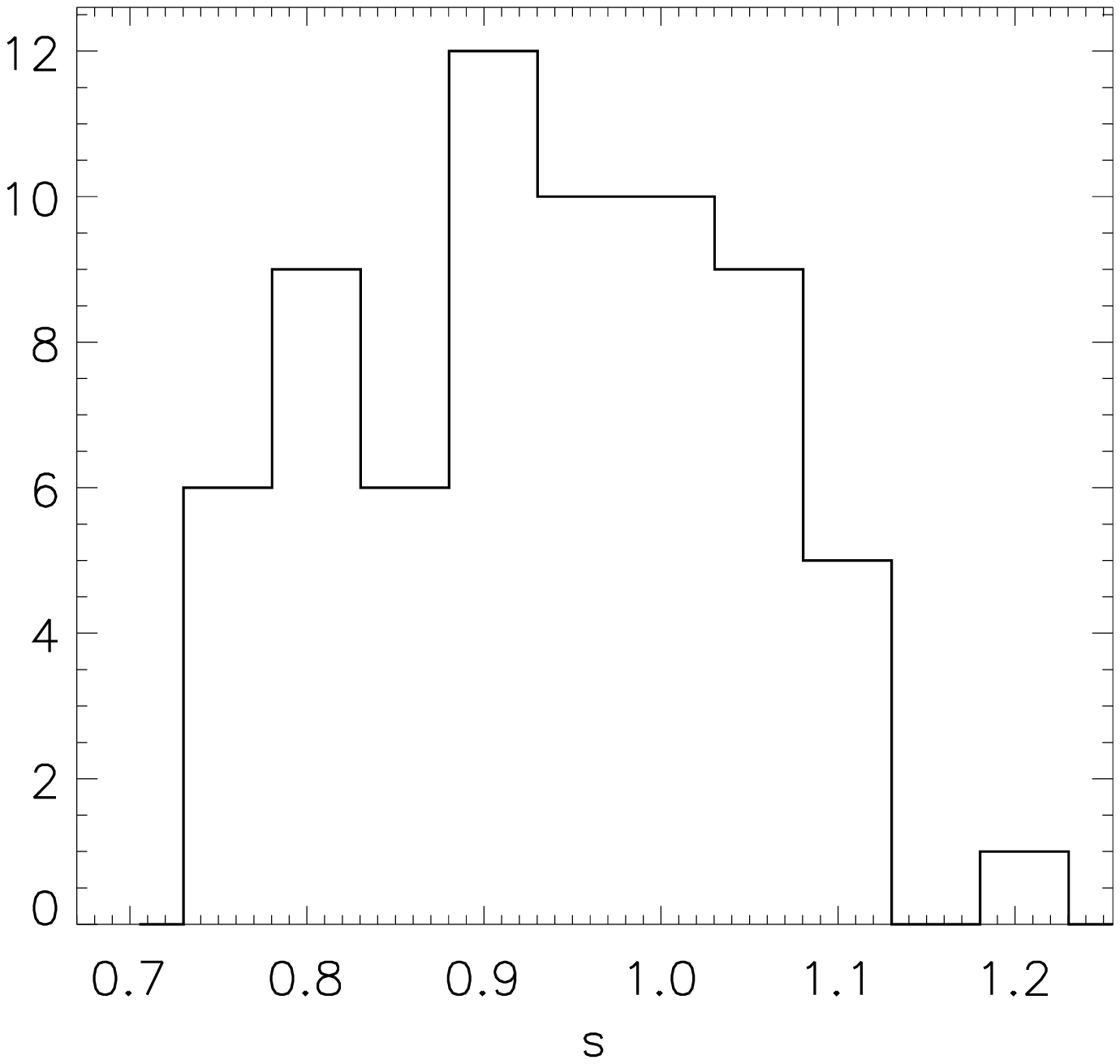}
\centering \includegraphics[width=8cm]{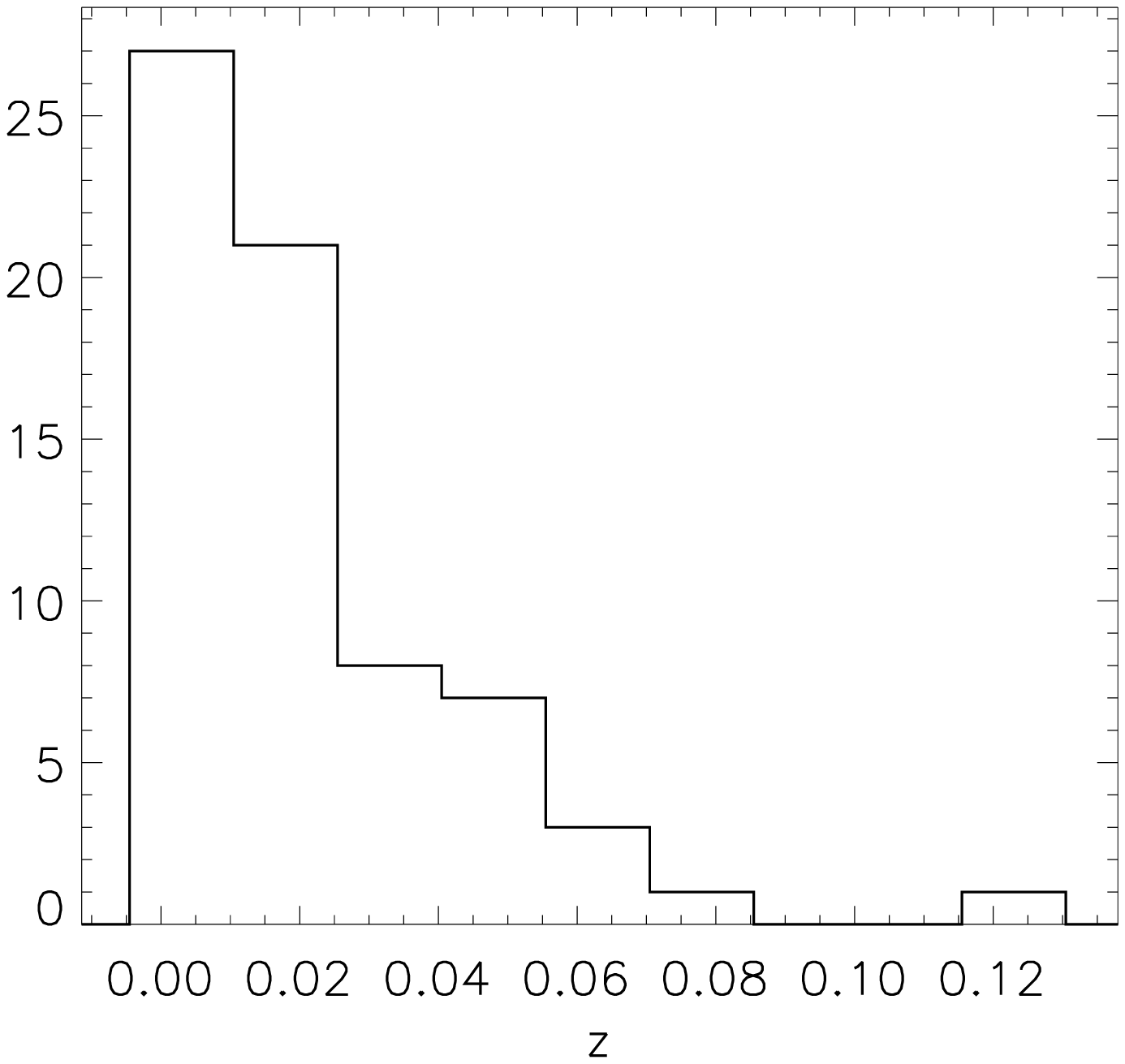}
\caption{Distribution of the $B$-band stretch factor,
  $s$, (top panel), distribution of the redshift 
(bottom panel) for the low extincted sample of 69 SNe.}
\label{histo}
\end{figure}

We present the analysis of the colour properties of a collection of 80
 well observed nearby SNe~Ia available from the literature.  This
 includes the Calan/Tololo data set \citep{1996AJ....112.2398H} and
 the CfA data set \citep{1998AJ....116.1009R}, analyzed in
 \citet{2003A&A...404..901N}, and the more recent data set by
 \citet{2006AJ....131..527J}, together with some single very well
 observed nearby supernovae (SN~2001el \citep{2003AJ....125..166K},
 SN~2000ca and SN~2001ba \citep{2004AJ....127.1664K}; SN~2001cn
 \citep{2004AJ....128.3034K}; SN~2001V \citep{2003A&A...397..115V};
 SN~2002er, \citep{2004MNRAS.355..178P}; SN~2003du
 \citep{2007A&A...469..645S}; SN~2005cf \citep{2007MNRAS.tmp..161P}).
 Sub-luminous Type Ia SNe, 1991bg-like, were not included in the sample
 nor supernovae with poor 
$B$-band sampling around lightcurve maximum. Out of these 80 SNe, a
subset with moderate colour excess, $E(B-V) < 0.25$
 mag\footnote{In section 6 we will relax this constrain, adding the 
remaining 11
 SNe, and study the impact on the derived reddening law}, was selected
to study the intrinsic colours of Type Ia supernovae. 
The complete list of the SNe used is reported in Table~\ref{listSNe}
together with the observed filter data available, redshift and
fitted $B$-band stretch factor ($s$) \citep{1997ApJ...483..565P,2001ApJ...558..359G}.
Figure~\ref{histo} shows the distribution of stretch factors and redshifts
for the whole sample.  \addtocounter{table}{1}

The data have been $K$-corrected to the rest frame bandpass, following
\citet{1996PASP..108..190K}, assuming Bessell filter transmission
curves \citep{1990PASP..102.1181B}. The spectral template used for
computing the $K$-corrections was obtained by
\citet{2007ApJ...663.1187H} by averaging about 600
spectra of SNe~Ia at different epochs.  The sample used for this
analysis includes nearby supernovae with redshifts up to
$\sim$0.1, with most of the SNe at $z < 0.06$ (see
Fig.~\ref{histo}).  Our ability to correct for extinction, i.e. to
compute the colour excess $E(B-V)$, depends on our knowledge of the
intrinsic colours of SNe~Ia. Thus, an iterative procedure has been
adopted until the procedure converged, as explained in
the next section.

\section{The colour-stretch relation and colour time evolution}
\label{sec:average}

The correlation between colours and lightcurve shape parameter 
is a known property of SNe~Ia. \citet{1999AJ....118.1766P} showed
the dependence of $B-V$ and $V-I$ at lightcurve maximum on the shape 
parameter. We use the larger data set available to extend the 
analysis to all epochs up to 60 days after lightcurve maximum. We also
expand the wavelength range to also include the $U$-band. 
As in \citet{2003A&A...404..901N}, the SN colours are derived from the
data points, without any interpolation.
For this to be possible, the SN should be observed in
at least two different bands at the same epoch, which is often the
case. 
Thus, the $B$-band lightcurve fit is performed exclusively with
the aim of determining the time of maximum and the stretch factor. 
We consider the rest frame time since $B$-band maximum, $t$, and the stretch 
factor, $s$, as two independent variables. We also introduce the colour 
excess $E(B-V)$ for each SN as independent variable. Each
colour $U-B$, $B-V$, $V-R$ and $R-I$ is fitted with a parametrised function:

\begin{equation}
X-Y=b_{XY}(t)+a_{XY}(t) \cdot (s-1)+c_{XY} \cdot E(B-V)
\label{eq:col}
\end{equation}

\noindent where $X$ and $Y$ are arbitrary filters, $a(t)$ and $b(t)$
are determined as least squares cubic spline fit of 6 knots each, distributed in
the time range between -10 and +60 days with respect to the time of
maximum light.  The knot positions are chosen to minimize
the number of knots while yielding a good fit to the data.
Note that, in this parametrisation, the $b$ functions correspond to
the colour curves for a stretch factor $s=1$ SN, while the $a$
functions give the dependence of the colours on stretch. Different
dependences on stretch have been investigated, e.g.~$(1/s^3-1)$ used by
\citet{2003ApJ...598..102K}. However, the simplest choice of
Eq.~\ref{eq:col} describes the current data best. The
$c_{XY}$ parameters are fitted for
each of the colours without assuming any specific
wavelength dependence of the dimming of Type Ia SNe.

\noindent An interactive procedure was applied as follows:

\begin{enumerate}

\item Fit the $B$-band lightcurve time of maximum
and the stretch parameter, following \citet{2001ApJ...558..359G}
\item Compute the colour $U-B$, $B-V$, $V-R$ and $R-I$ for each SN at 
each observed epoch
\item Correct the observed colours for Milky-Way extinction, assuming
  the extinction law \citep{1989ApJ...345..245C}(CCM), as modified by
  \citet{1994ApJ...422..158O}. The corrections are computed by
  reddening the SNe~Ia spectral template at different epochs, and
  integrating in all band-passes.
\item Compute $K$-corrections for each SN using the spectral template
  corrected by the colour-stretch relation determined in the previous
  iteration. In the first iteration,
  $K$-corrections where computed without using any colour-stretch
  relation, i.e. as if all SNe had $s=1$.
\item Compute the colour excess $E(B-V)$ as weighted average excess for all epochs 
from the $B-V$ colour curve for each supernova determined in the
previous iteration. The inverse square of the uncertainty on the individual colour
measurements, $1/\sigma_i^2$, is used as weights in the average, and the
uncertainty on $E(B-V)$ is determined
as $1/\sqrt{\sum 1/\sigma_i^2}$. For the first iteration, an initial colour
excess $E(B-V)$ for each SN was estimated following the method
described by \citet{1999AJ....118.1766P}.
\item Fit the function in Eq.~\ref{eq:col} to the data separately for each colour.
\item Modify the spectral template for the average colour-stretch relation.

\end{enumerate}

\noindent The spectral template colours are modified by using a cubic
spline interpolation of the ratio between the synthetic photometry and
the new photometry at the effective wavelengths, for the $U$, $B$, $V$, $R$
and $I$-band.  These are determined from the fitted colour curves as follows:

\begin{equation}
\begin{array}{l}
U=(U-B)+B'\\
B=B'\\
V=B'-(B-V)\\
R=V-(V-R)\\
I=R-(R-I)\\
\end{array}
\label{eq:warp}
\end{equation}

\noindent where $B'$ indicates the $B$-band lightcurve by
\citet{2001ApJ...558..359G}, and the colours between brackets are the
one fitted in the previous iteration.

The steps 1-7 were repeated until the difference 
in the colour-curves between consecutive iterations changed by  
less than 0.5\%. Only three iterations were needed to reach such precision.

\begin{figure*}[htb]
\centering \includegraphics[width=12cm]{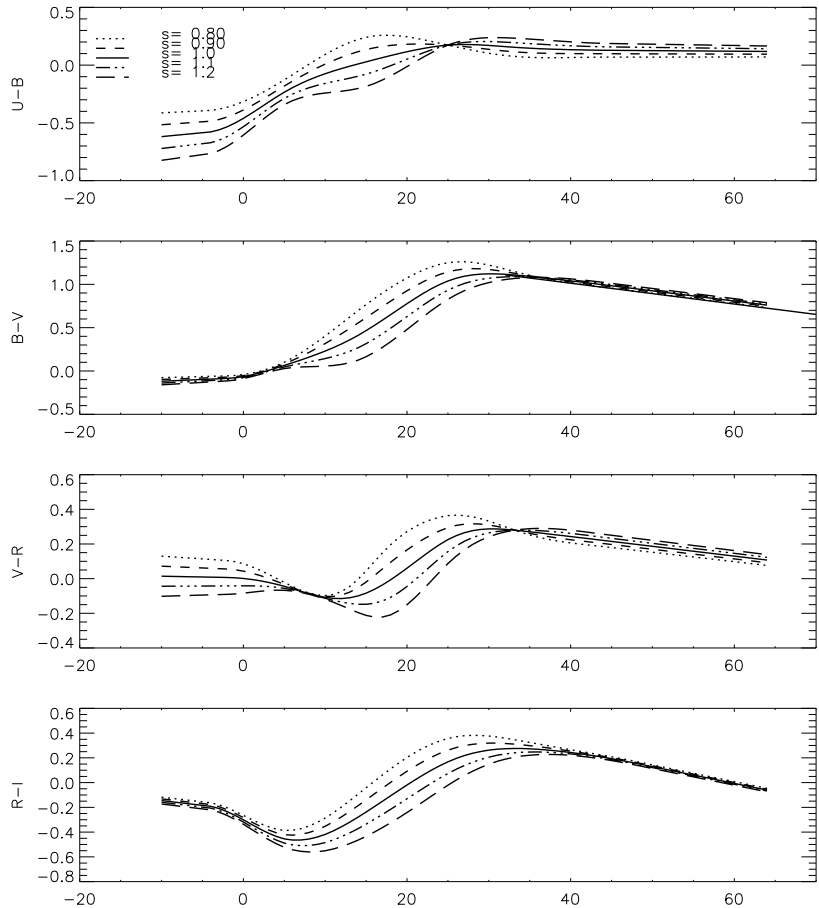}
\caption{From top to bottom, $U-B$, $B-V$, $V-R$ and $R-I$ time
evolution for different stretch values: $s=0.8,0.9,1.0,1.1,1.2$.
The straight line at late time in the $B-V$ evolution is the Lira 
line. The abscissa is the restframe epoch since the $B$-band 
maximum light.}  
\label{models}
\end{figure*}

\begin{figure*}[htb]
\centering \includegraphics[width=12cm]{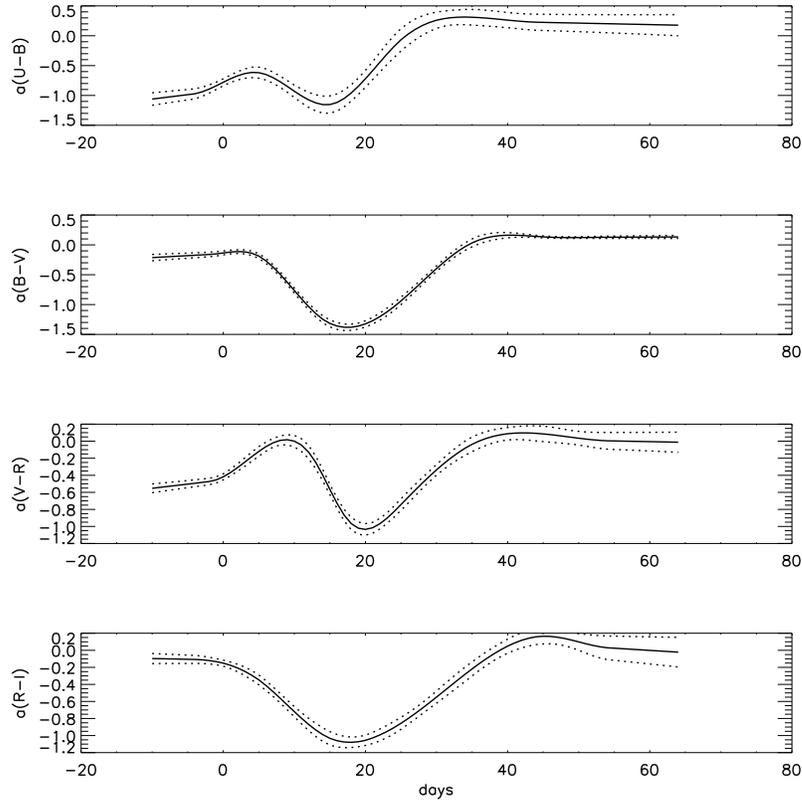}
\caption{From top to bottom, the $a$ parameter time evolution for
$U-B$, $B-V$, $V-R$ and $R-I$ (solid line), and its uncertainty
(dotted line). The minimum in the curves from top to bottom 
occurs at day 15, 17, 20 and 20.}
\label{modela}
\end{figure*}

Tables~\ref{table:abUB} -~\ref{table:abRI} report the values
for $a$ and $b$ for each colour as a function of time.
Figure~\ref{models} shows the time evolution of $U-B$, $B-V$, $V-R$
and $R-I$ for different values of the stretch factor,
$s=0.8,0.9,1.0,1.1,1.2$. Figure~\ref{modela} gives a comparison of
the $a$ parameter fitted for each of the colours. For $U-B$ the
dependence on stretch is substantial already before maximum light, and
decreases with time (see discussion in Section~\ref{sec:discussion}).  
In all other cases, the dependence on
stretch peaks between 15 and 20 days after $B$-band maximum.
This coincides with the epoch at which the photosphere recedes into the 
core of the supernova, and absorption features from iron group elements 
appear in the spectrum. \citet{2007ApJ...656..661K} show that 
the colour evolution depends on the ionization evolution of the 
iron group elements, which is faster for dimmer supernovae. 
Their model dependence is in good agreement with the 
results of this work.

Note that the data points were considered to be uncorrelated at
different epochs when performing the fit. Thus, even if some SN
lightcurve are better sampled than others, all points have the same
weight in the fit. As this is only a crude approximation, we take this 
into account in the determination of the uncertainties. For this
reason, the uncertainties on the colour curves reported in 
Tables~\ref{table:abUB} -~\ref{table:abRI}, are determined by the use of
Monte-Carlo simulations, as explained in Section~\ref{sec:MC}.

\begin{figure*}[htb]
\centering 
\includegraphics[width=8cm]{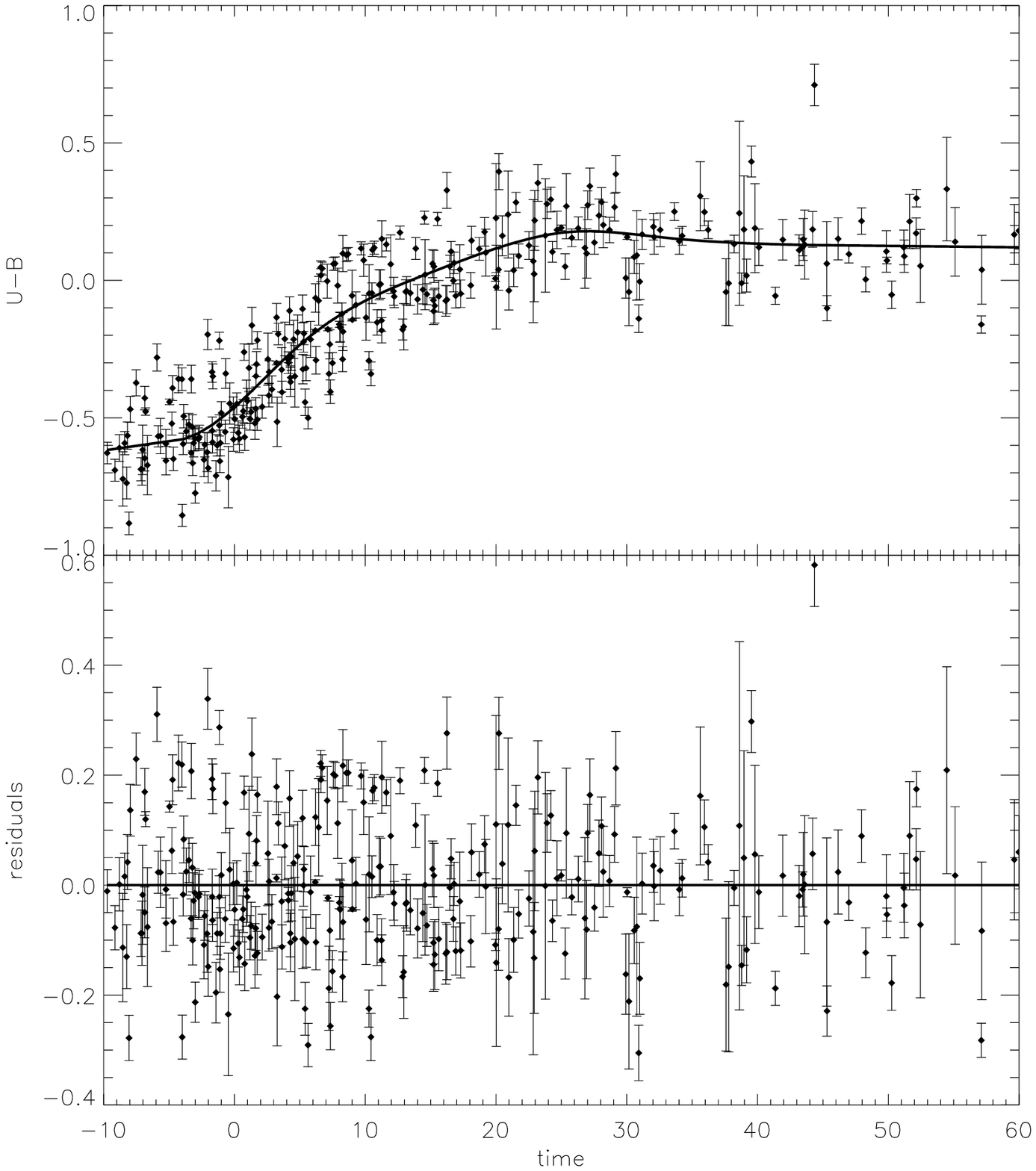}
\includegraphics[width=8cm]{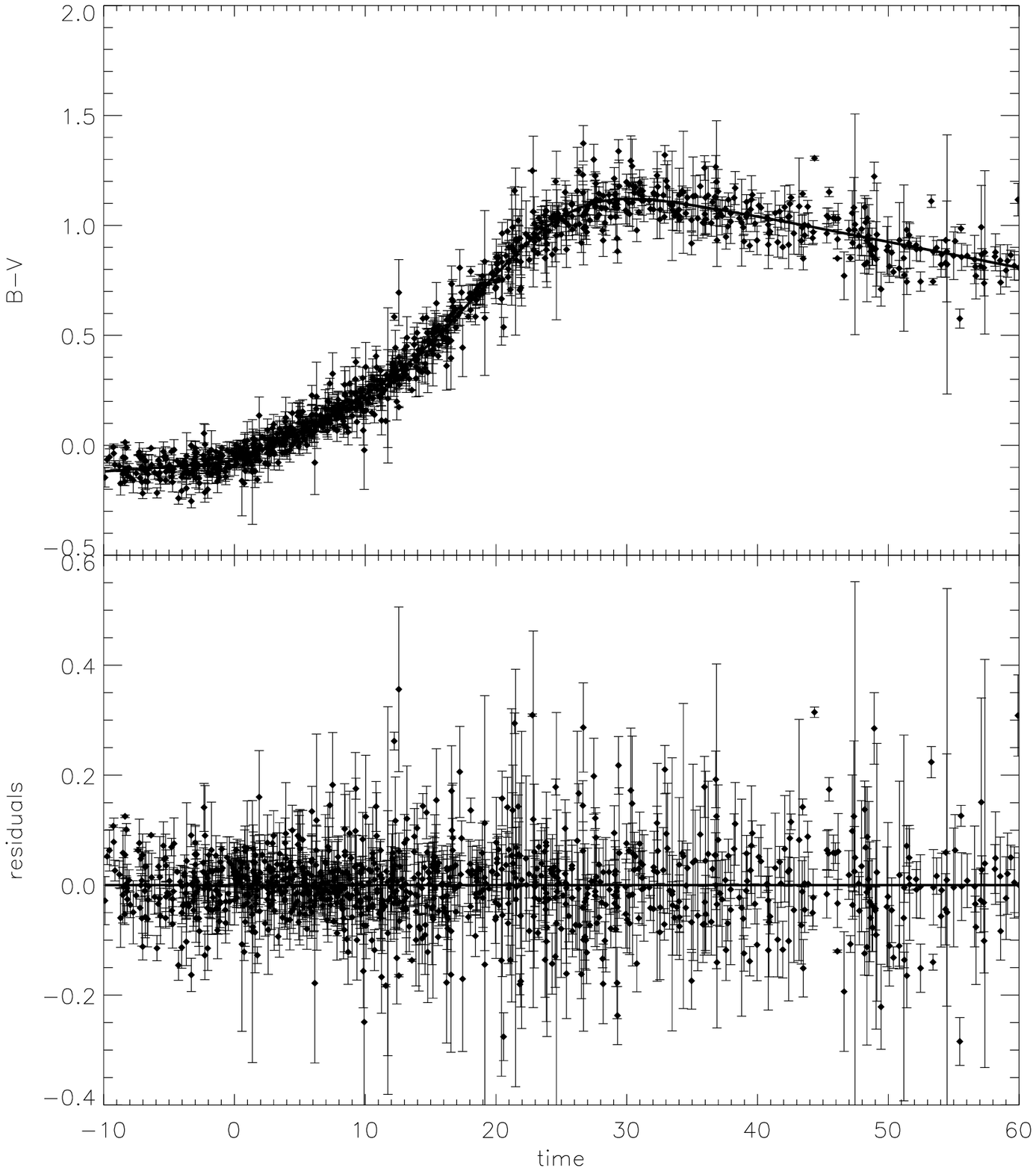}
\includegraphics[width=8cm]{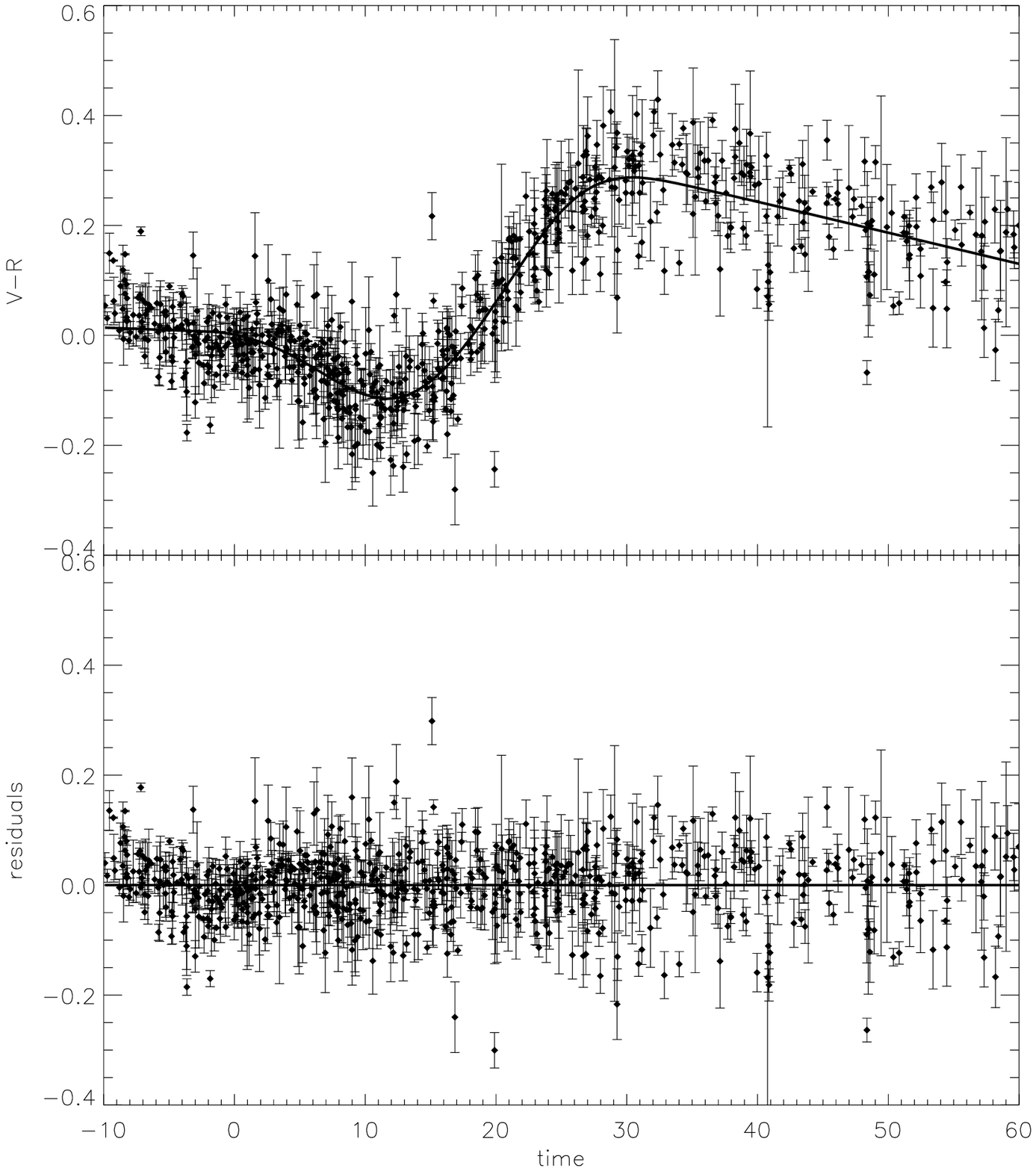}
\includegraphics[width=8cm]{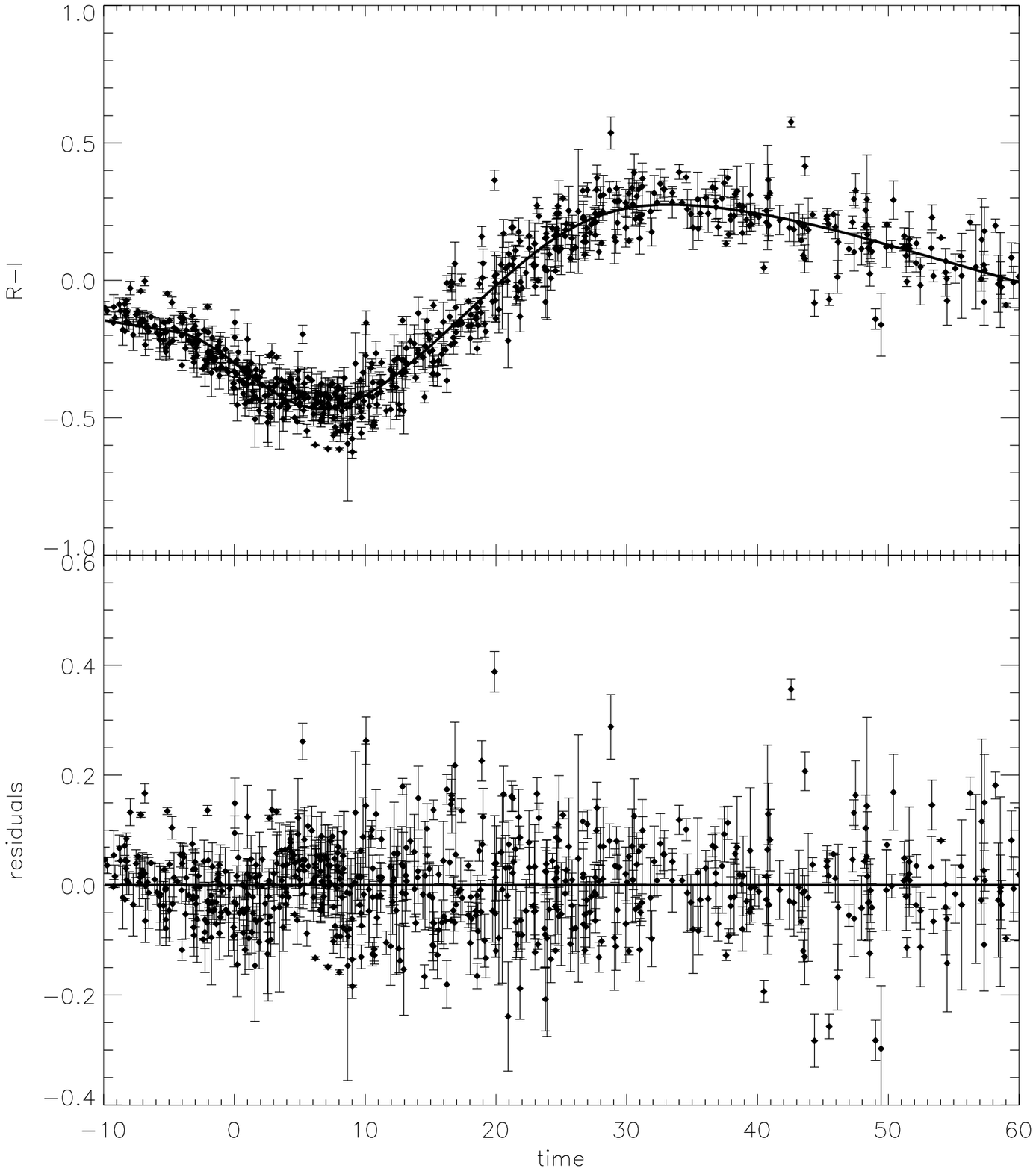}
\caption{From top to bottom, left to right, $U-B$, $B-V$, $V-R$ and $R-I$ time
evolution compared with the data corrected for Milky Way extinction 
as well as for the colour-stretch relation and host galaxy extinction 
following Eq.~\ref{eq:col}. }
\label{data}
\end{figure*}

\section{Intrinsic colour dispersion}
\label{sec:intrinsic}

The found colour-stretch relation contributes to decreasing the
dispersion along the colour curve for all colours, once it is applied
to the data. However, the remaining dispersion is still too large to be
explained solely by statistical fluctuations based on the measurement errors. 
This forces  us to conclude that there is remaining intrinsic
dispersion in the SN colours.  We follow the same procedure as in
\citet{2003A&A...404..901N} in order to estimate the intrinsic dispersion for
each colour as a function of time. The data, corrected for the
colour-stretch relation found in the previous section, are divided in
time bins. For each time bin, we compute the weighted average of the
residual to the colour curve (for a $s=1$ supernova), and the sample
weighted standard deviation, as the square root of the weighted second
moment (see \citet{2003A&A...404..901N}). Table~\ref{intrdisp} gives the
results of this analysis.

The weighted standard deviation can be taken as an estimate of the
intrinsic dispersion in each time bin, $\Delta$, however, as already
pointed out in \citet{2003A&A...404..901N}, this is an overestimate since part
of the scatter is due to measurement uncertainty only. In order to
disentangle the two contributions, we run a Monte Carlo simulation to
generate synthetic data sets, with dispersion given by the quoted
measurement uncertainties only. Thus, we compute the weighed standard
deviation on the simulated data sets, $\delta$. Finally we run a
hypothesis test, to compare the dispersion measured on the real data
to those measured on the synthetic data sets for each colours. 

\noindent We set a level of significance $\alpha=0.01$ for rejecting the 
null hypothesis, i.e. the probability 
that $\Delta \ne \delta$ under the assumption that the null hypothesis is 
correct is assumed equal to 1\% \citep{2003A&A...404..901N,Cowan1998}.
For all cases for which the null hypothesis is rejected, we compute the quantity:

\begin{equation}
\Delta^{\rm corr}=\sqrt{\Delta^2-\delta^2}
\label{eq:intr}
\end{equation}

that we take as an estimate of the intrinsic dispersion, and a lower
limit on this value is set at a 99\% confidence limit. 
We found no cases compatible with {\it no intrinsic dispersion}.
(see Table~\ref{intrdisp} ). 

\begin{table}[!htb]\caption[table data]{Results of the analysis of all SNe. 
First column: central value in days for each time bin; $N_k$ is the
number of points for each bin; $\Delta_{XY}$ is the intrinsic dispersion
computed as weighted standard deviation; $\Delta_{XY}^{\rm corr}$ is
the corrected intrinsic dispersion, computed as in Eq.~\ref{eq:intr},
and in the last column is the estimated lower limit at 99\% C.L. (see
text); }
\begin{center}
\begin{tabular}{rrcll}
\hline
\hline
day & $N_k$ &   $\Delta_{UB}$ & $\Delta_{UB}^{\rm corr}$ & L.L. \\
\hline
   -10  & 22   &    0.13 $\pm$   0.03  &    0.13  &    0.09 \\
     0  & 83   &    0.10 $\pm$   0.03  &    0.10  &    0.06 \\
    10  & 67   &    0.06 $\pm$   0.05  &    0.06  &    0.03 \\
    20  & 44   &    0.10 $\pm$   0.02  &    0.09  &    0.06 \\
    30  & 29   &    0.06 $\pm$   0.07  &    0.06  &    0.04 \\
    40  & 20   &    0.07 $\pm$   0.02  &    0.06  &    0.05 \\
\hline
day & $N_k$ &   $\Delta_{BV}$ & $\Delta_{BV}^{\rm corr}$ & L.L. \\
\hline
    -5  & 77   &    0.05 $\pm$   0.01  &    0.05  &    0.03 \\
     0  & 180   &    0.03 $\pm$   0.01  &    0.03  &    0.02 \\
    10  & 228  &    0.08 $\pm$   0.01  &    0.08  &    0.04 \\
    18  & 165   &    0.05 $\pm$   0.01  &    0.05  &    0.02 \\
    30  & 137   &    0.10 $\pm$   0.01  &    0.09  &    0.05 \\
    40  & 74    &    0.06 $\pm$   0.01  &    0.06  &    0.03 \\
    50  & 66    &    0.12 $\pm$   0.04  &    0.12  &    0.06 \\
\hline
\hline
day & $N_k$ &   $\Delta_{VR}$ & $\Delta_{VR}^{\rm corr}$ & L.L. \\
\hline
    -5  & 64   &    0.04 $\pm$   0.01  &    0.04  &    0.02 \\
     0  & 120  &    0.04 $\pm$   0.01  &    0.04  &    0.02 \\
    10  & 131  &    0.03 $\pm$   0.01  &    0.03  &    0.02 \\
    15  & 65   &    0.08 $\pm$   0.01  &    0.08  &    0.04 \\
    20  & 82   &    0.03 $\pm$   0.01  &    0.03  &    0.01 \\
    30  & 81   &    0.03 $\pm$   0.05  &    0.03  &    0.02 \\
    40  & 58   &    0.05 $\pm$   0.01  &    0.05  &    0.03 \\
    50  & 47   &    0.03 $\pm$   0.01  &    0.03  &    0.02 \\
\hline
day & $N_k$ & $\Delta_{RI}$ & $\Delta_{RI}^{\rm corr}$ & L.L. \\
\hline
   -10  & 44    &    0.04 $\pm$   0.01  &    0.04  &    0.03 \\
     0  & 147   &    0.06 $\pm$   0.01  &    0.06  &    0.03 \\
    10  & 141   &    0.10 $\pm$   0.02  &    0.10  &    0.05 \\
    20  & 77    &    0.13 $\pm$   0.02  &    0.13  &    0.07 \\
    25  & 97    &    0.05 $\pm$   0.02  &    0.05  &    0.03 \\
    40  & 62    &    0.14 $\pm$   0.03  &    0.14  &    0.08 \\
    50  & 50    &    0.14 $\pm$   0.04  &    0.14  &    0.08 \\
\hline
\end{tabular}
\label{intrdisp}
\end{center}
\end{table}

\section{Supernova spectral template}
\label{sec:template}

The spectral template built by \citet{2007ApJ...663.1187H}, (``{\hsiao}''
in our plots), was used as the starting assumption in the
iterative method for this analysis. The template
was built using $\sim$600 spectra for $\sim$100 SNe~Ia where the
observed spectra were colour-corrected to match a single SED, 
before averaging
them into one uniform spectral template. The colours used were those in
\citet{2003ApJ...598..102K}, corresponding to a ``normal'' $s=1$ SNe~Ia.  
The {\hsiao} spectral
template is meant to be used for computing $K$-corrections after
correcting it for the observed colours. This has the advantage of
being independent of an assumed stretch-colour relation, but the
resulting uncertainties on the $K$-corrections, depend on the
availability of observed colours, and can be as large as 0.12 mag, as
showed in Fig.~9 of  \citet{2007ApJ...663.1187H}.  
By using the colour-stretch relation derived with our technique, 
the spectral template colours can be adjusted for each supernova lightcurve
shape,
before computing $K$-corrections. In our case, the uncertainties 
on $K$-corrections are
dominated by the intrinsic dispersion in the supernova colours.
Whether these correspond to an intrinsic variability of the spectral
features is beyond the scope of this work. We note, however, that
there are several indications that at least some of the characteristic
SN~Ia spectral features correlate well with lightcurve shape
parameters \citep{1995ApJ...455L.147N, 2005ApJ...623.1011B, 2007A&A...470..411G}.

Another interesting approach is the one in {\salt}
\citep{2007A&A...466...11G}. Instead of building spectral templates to
be used for computing $K$-corrections, an empirical model that
describes the time variation of the spectral energy distributions and
its dependence on a lightcurve shape parameter (corresponding to
stretch) is used. In this case, the SED is a function of epoch, wavelength and 
lightcurve shape parameter.
This ambitious model is trained on about the same data set as the
{\hsiao} templates. However, due to the larger amount of parameters
needed to describe the model, the uncertainties are quite large,
especially in the $U$ and $I$ part of the spectrum, where fewer measurements
are available. 

\begin{figure*}[htb]
\centering \includegraphics[width=9cm]{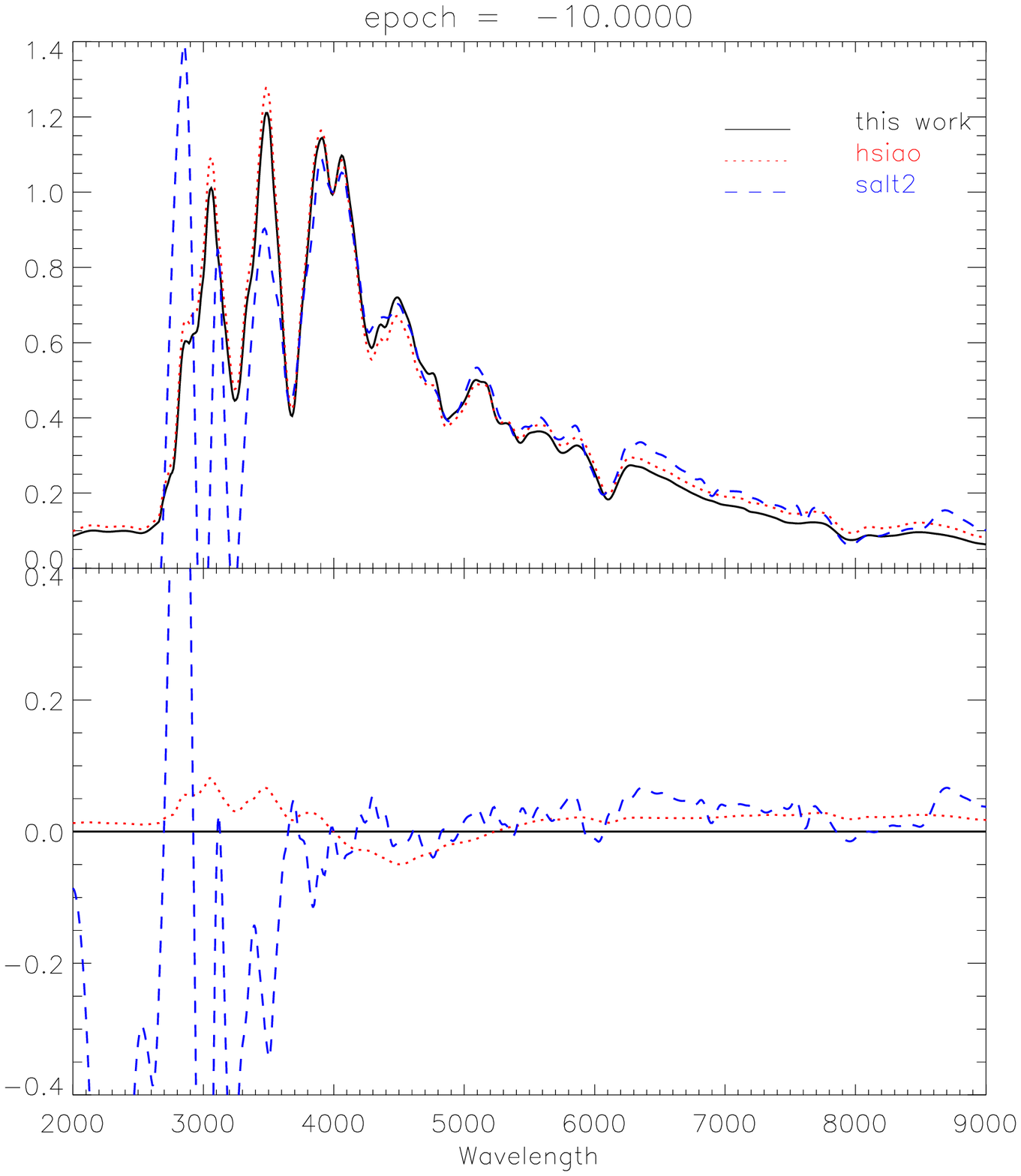}
\centering \includegraphics[width=9cm]{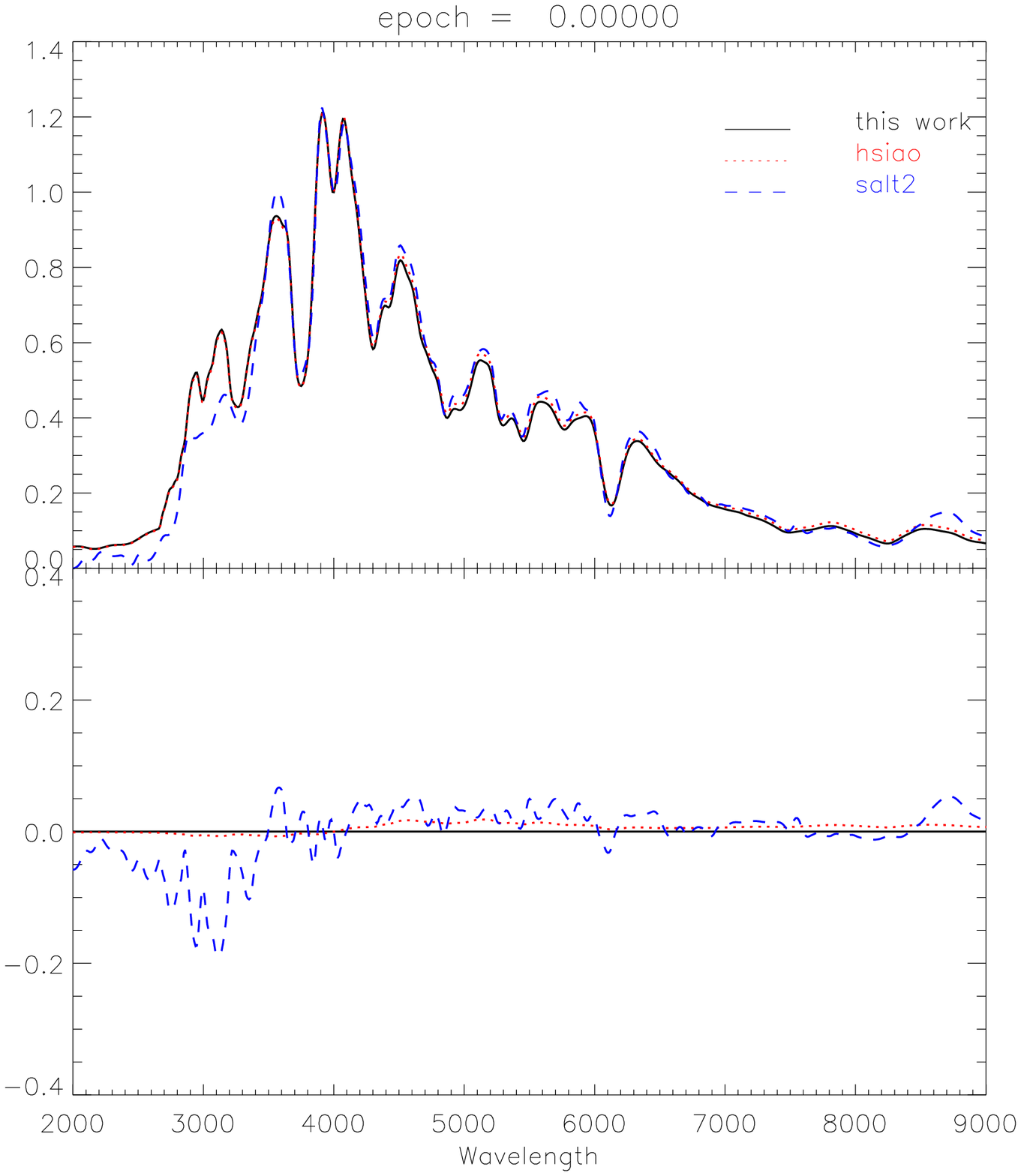}
\centering \includegraphics[width=9cm]{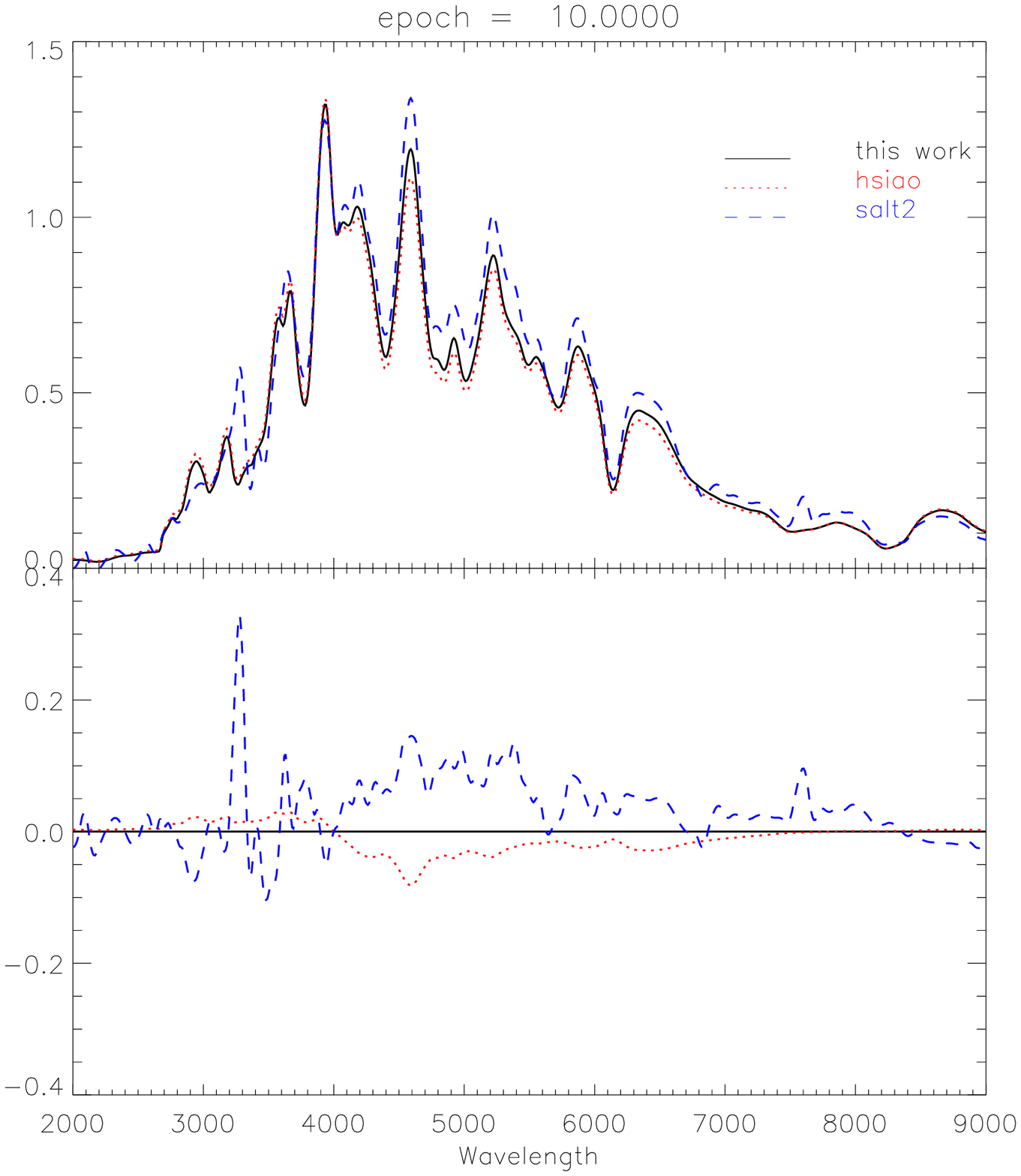}
\centering \includegraphics[width=9cm]{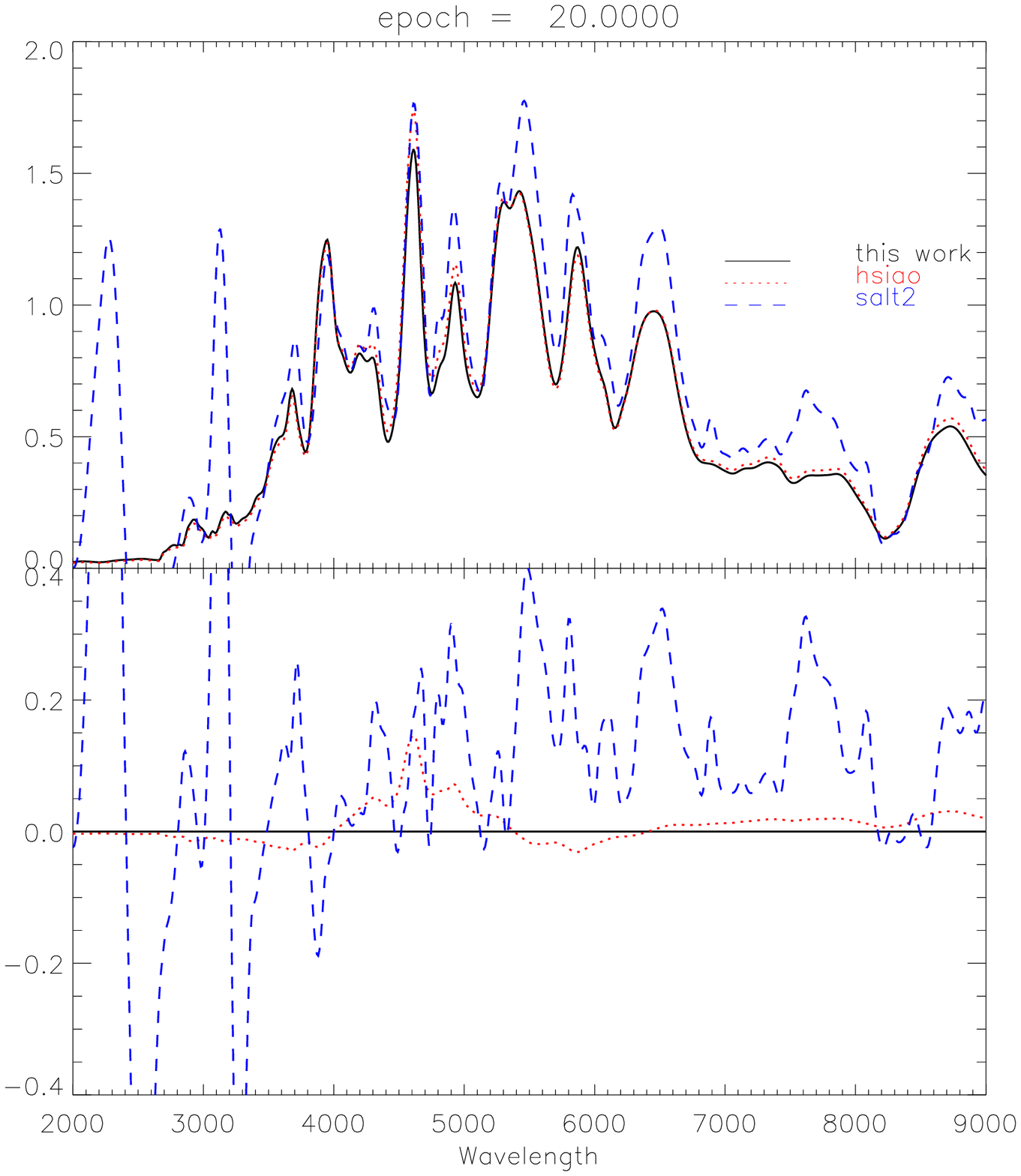}
\caption{Comparison of the spectral templates derived in this work for $s=1$ (black solid line),
the spectral template by  \citet{2007A&A...466...11G} (blue dashed line)
and the one by \citet{2007ApJ...663.1187H} (red dashed-dotted line) for epochs=-10,0,10,20 
referred to the time of maximum.}
\label{spectra}
\end{figure*}

A comparison between the spectral
templates for a $s=1$ supernova at some epochs from maximum light
(day=-10,0,10,20) is shown in Fig.~\ref{spectra}.  Differences
in our colour corrected spectral template and the original {\hsiao}
template are small, up to 2\%, and even smaller around maximum,
about 0.5\%. This differences
originate from different data sets used to determine intrinsic colours.
The {\hsiao}-template uses the colours in 
\citet{2003ApJ...598..102K}, based on a smaller sample than
the one used here. We note in particular, the $U-B$ colour that was
derived based on measurements of only 5 SNe~Ia at maximum.  
When comparing our spectral template with the {\salt} template we
find large differences already for $s=1$.  
We note, however, that the differences are mainly
in the $U$ part of the spectra where large uncertainties of the 
{\salt} templates are reported by the authors.

To further compare the spectral templates, we have computed
$K$-corrections using {\salt} templates, {\hsiao} templates and our
templates for different values of the redshift.  The results of the
comparison for $s=1$ are reported in Table~\ref{comp_kcorr}. The
larger discrepancy is found for $z=0.6$, where the mismatch of the
observed and restframe filter is largest. This is in agreement with
the results found by \citet{2007ApJ...663.1187H}, suggesting an
uncertainty of $\sim$ 0.04 in $K$-corrections for this
redshift.

\begin{table}[htb]
\caption{Comparison at different redshifts between the $K$-corrections computed using the
various spectral templates for $s=1$. The corrections are computed from
the band listed in the first column to restframe $B$-band.
The mean and the r.m.s. are computed over time on the residuals to the
$K$-corrections computed using our templates. 
The largest dispersion is noted at $z=0.6$ where the mismatch between observed and 
restframe filters is maximum, as also noted by \citet{2007ApJ...663.1187H}}
\begin{center}
\begin{tabular}{ccrr}
\hline\hline
Band     & $z$  & mean   &  r.m.s. \\
\hline
\hsiao  & & & \\
\hline
B   &   0.1 &  0.008 &  0.020  \\
V   &   0.3 &  0.007 &  0.015  \\
R   &   0.6 &  0.012 &  0.031  \\
I   &   0.8 & -0.011 &  0.028  \\
I   &   0.9 &  0.005 &  0.010  \\
\hline
\salt & & & \\
\hline
B   &   0.1 &  0.023 &  0.017  \\
V   &   0.3 &  0.021 &  0.015  \\
R   &   0.6 &  0.041 &  0.040  \\
I   &   0.8 & -0.040 &  0.026  \\
I   &   0.9 &  0.016 &  0.013  \\
\hline
\end{tabular}
\label{comp_kcorr}
\end{center}
\end{table}

\begin{figure}[htb]
\centering 
\includegraphics[width=8cm]{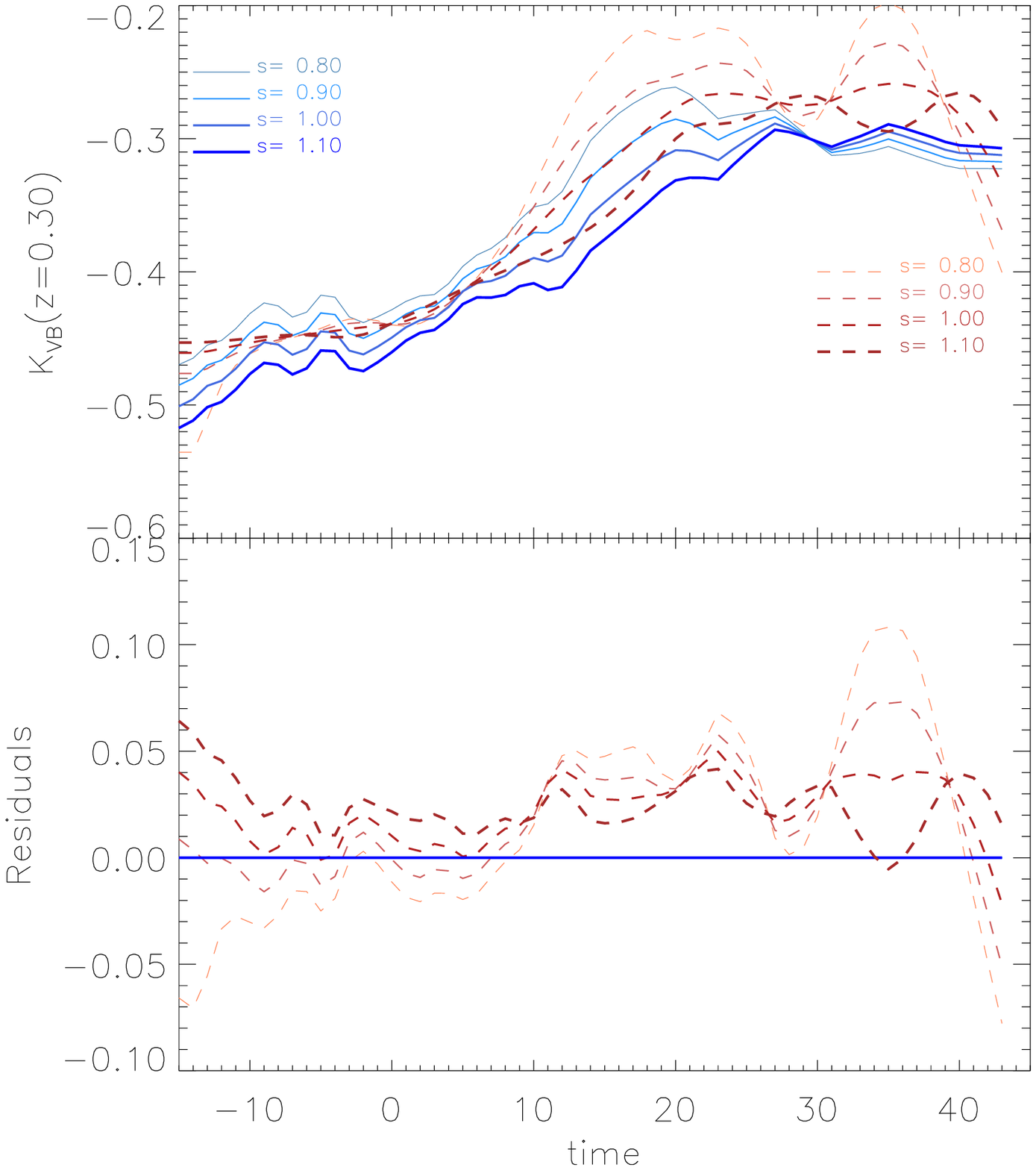}
\includegraphics[width=8cm]{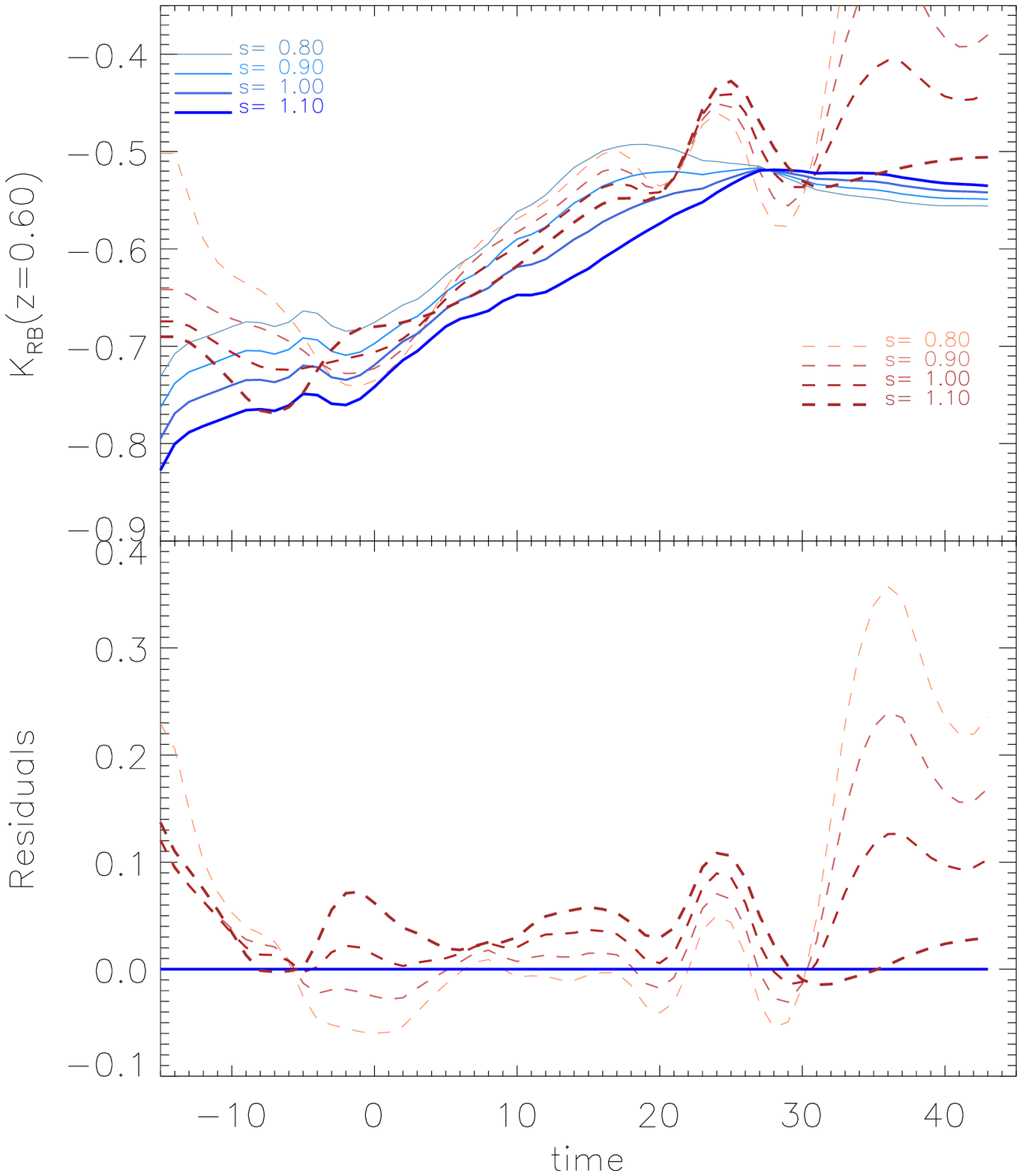}
\caption{$K$-corrections to restframe $B$-band at redshift $z=0.3$ (top panel)
and $z=0.6$ (bottom panel) using our templates (solid lines) and {\salt} templates
(dashed lines).}
\label{kcorr_str}
\end{figure}

A comparison between the dependence of $K$-corrections on stretch, 
is only possible between {\salt} and our templates. 
Fig.~\ref{kcorr_str} shows the $K$-corrections to restframe $B$-band for 
two different redshifts $z=0.3$ and $z=0.6$ computed using the two sets of 
templates. The r.m.s. on the difference is reported in Table~\ref{kcorr_str_table}. 

All the differences found demostrate that by using the uncorrected
spectral template, it is possible to introduce systematic
uncertainties as large as 0.04 mag, limiting our ability to use
SNe~Ia for precision cosmology. 

\begin{table}[htb]
\caption{Comparison between the dependence of $K$-corrections on stretch for 
two different redshifts, for {\salt} and our templates.
The mean and the r.m.s. are computed over time on the residuals to our templates. 
Once again the largest dispersion is noted at $z=0.6$ where the mismatch between 
observed and restframe filters is maximum, and for more extreme stretch values.}
\begin{center}
\begin{tabular}{llcrr}
\hline\hline
  $z$  &  $s$  &mean   &  r.m.s. \\
\hline
 0.3 &    0.8 &  0.014 &  0.045  \\
 0.3 &    0.9 &  0.019 &  0.028  \\
 0.3 &    1.0 &  0.022 &  0.015  \\
 0.3 &    1.1 &  0.023 &  0.013  \\
\hline
 0.6 &    0.8 &  0.061 &  0.123  \\
 0.6 &    0.9 &  0.049 &  0.078  \\
 0.6 &    1.0 &  0.042 &  0.041  \\
 0.6 &    1.1 &  0.036 &  0.033  \\
\hline
\end{tabular}
\label{kcorr_str_table}
\end{center}
\end{table}

\begin{figure}[htb]
\centering \includegraphics[width=8cm]{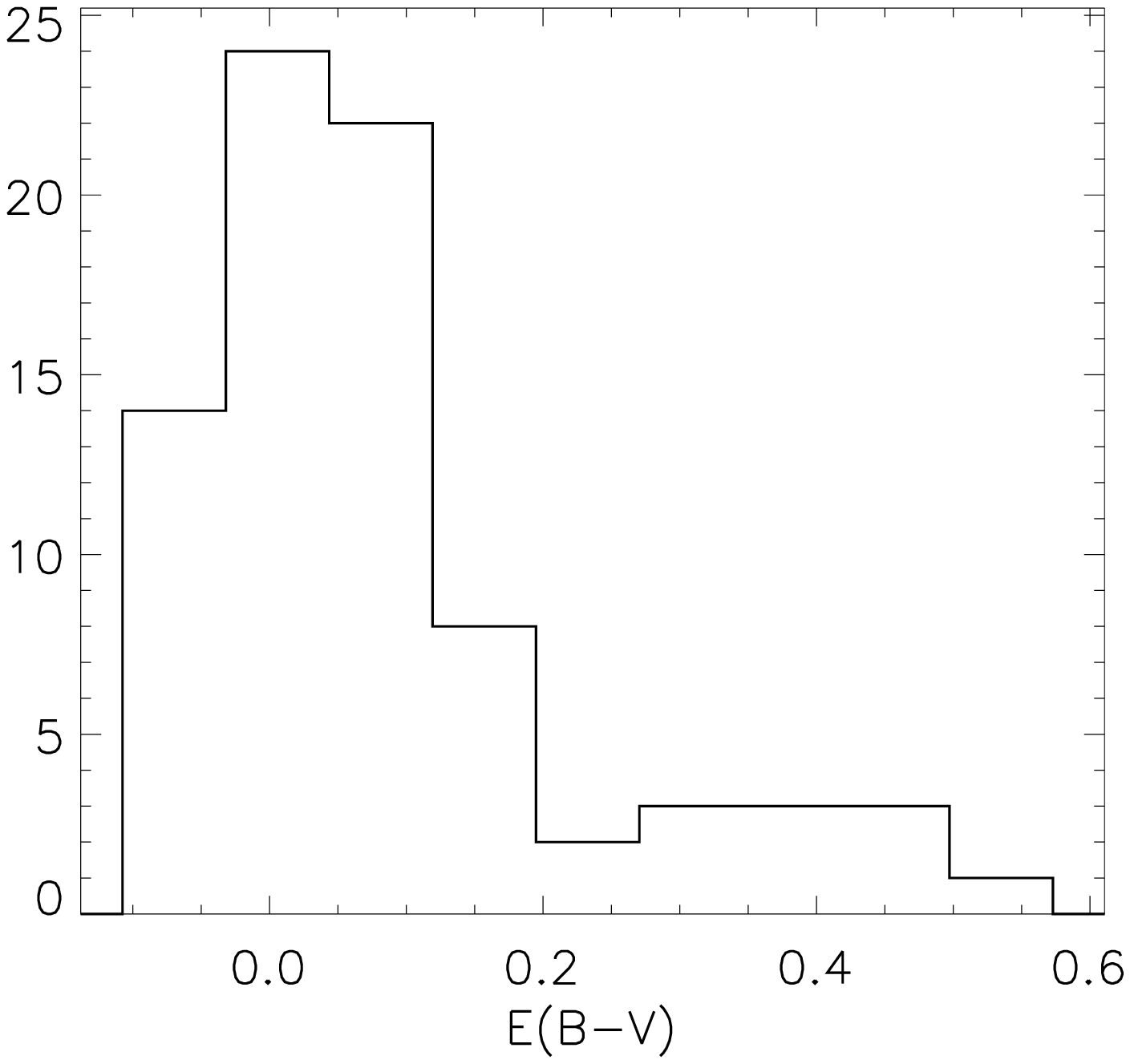}
\caption{Distribution of colour excess $E(B-V)$ for the whole sample including
 80 SNe.}
\label{histo2}
\end{figure}

\section{The reddening law from SNe~Ia}
\label{sec:extlaw}

Corrections due to the wavelength dependence of the dimming of the
supernova light is one of the largest systematic uncertainties in
supernova cosmology. While the empirical evidence of the 
dependence of peak magnitudes on SN colours is very clear, 
there is no consensus on how to disentangle the contribution 
from reddening due to dust in the host galaxy and 
 intrinsic SN colour variations. Moreover, most published
results assume the dust in the host galaxy to have similar properties
as (the average) Milky Way dust, implying a total to selective extinction
parameter $R_V = 3.1$. Studies of dust properties in small sets of
distant galaxies indicate mean values compatible with this value, even though  
a significant range of values may be present
\citep{2007arXiv0711.4267O,1994MNRAS.271..833G,2007A&A...461..103P}. 
Ideally, cosmological distance estimates
using SNe~Ia should include extinction corrections for each particular
line of sight, including the specific dust properties of each host
galaxy. However, as shown in \citep{2008arXiv0801.2484N},
the variations of $R_V$ between lines of sight are much less problematic
for cosmological applications than a potential bias in the adopted mean value.

Intriguingly, there are a large number of examples of measurements of
the mean $R_V$ from SNe~Ia, based on different methods, that disagree
with the universally assumed value: a significantely smaller value of
the mean $R_V$ is found.  Some examples are $R_V
\sim 1.8$ \citep{2000ApJ...539..658K}, $R_V = 2.55 \pm 0.30$
\citep{1996ApJ...473..588R}, $R_V = 2.6 \pm 0.4$
\citep{1999AJ....118.1766P}, $R_V = 1.09$ \citep{1998A&A...331..815T},
$R_V =2.5$ \citep{2004MNRAS.349.1344A}, $\beta=1.77 \pm 0.16$
corresponding to $R_V= 0.77 \pm 0.16$
\citep{2006A&A...447...31A,2007A&A...466...11G}.

\begin{figure}[htb]
\centering 
\includegraphics[width=8cm]{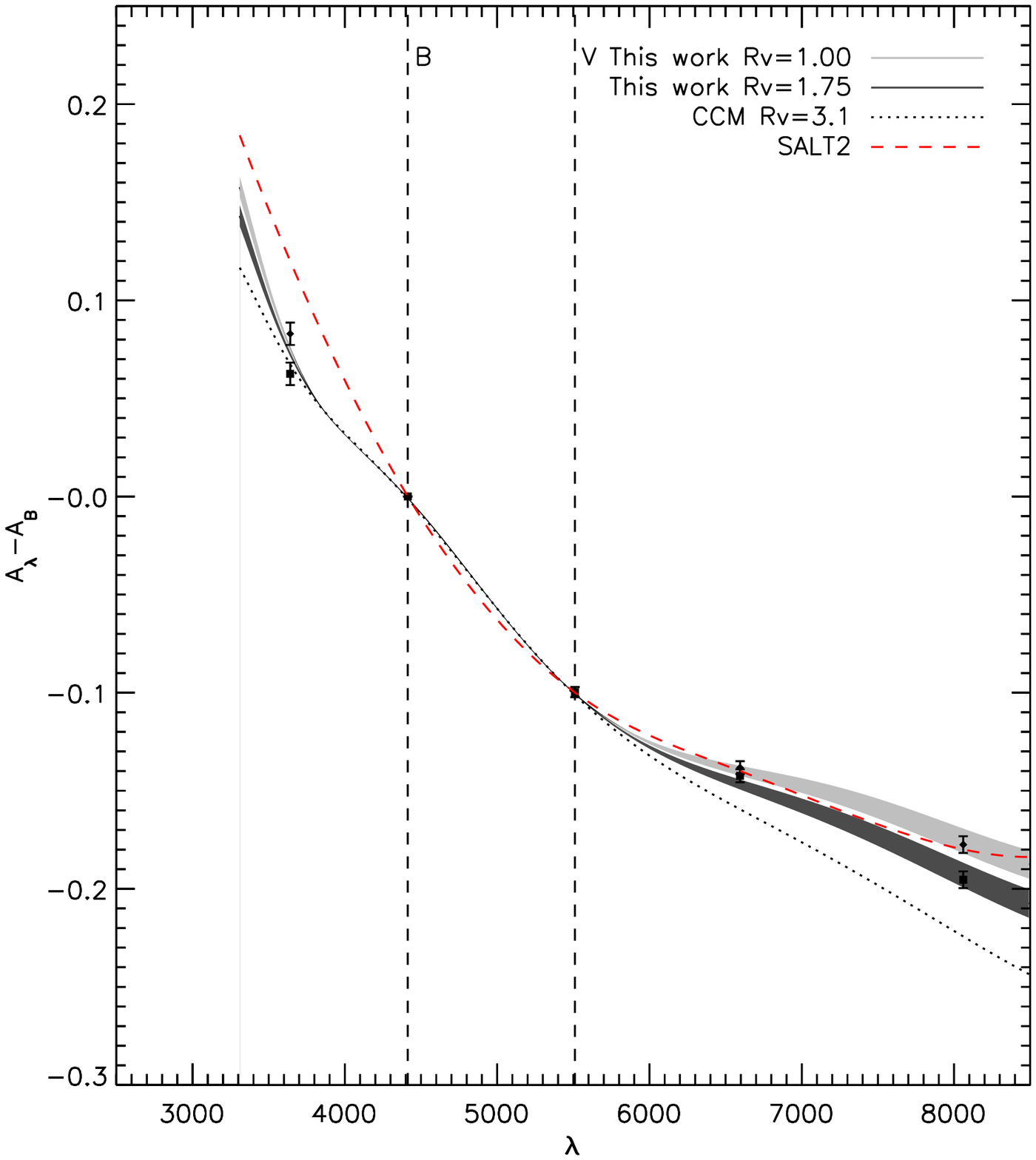}
\caption{The extinction law determined using the $c$ parameters in our model, 
compared with the one from \citet{1989ApJ...345..245C} with $R_V=1.01$
(light grey solid line) and
$R_V=3.1$ (dotted line). Also plot is the result by {\salt} (dashed red line).
The CCM law with $R_V=1.01$ give the best fit to our data (solid points). 
The lighted shaded region represent the variability on the $R_V$ parameter 
obtained by running the Monte Carlo simulation.
The darker shaded region is obtained by adding 11 highly extincted SNe, and 
the CCM law with $R_V=1.75$ give the best fit to our data (solid squares).
A value of $E(B-V)=0.1$ was assumed for all the curves in the plot.}
\label{mycardelli}
\end{figure}

Using the fitted $c$ parameters in Eq.~\ref{eq:col}
the reddening law that best describes the data can be determined. In 
fitting the $c$ parameter for each colour,
$c_{UB},c_{BV},c_{VR},c_{RI}$, we minimize the part of the
colour dispersion that depends on reddening, without assuming any
value for $R_V$.  The four $c$ parameters, one for each
colour fitted, can then be used to estimate the parameter $R_V$:

\begin{equation}
\begin{array}{c}
c_{UB}=R_U-R_B\\
c_{BV}=R_B-R_V\\
c_{VR}=R_V-R_R\\
c_{RI}=R_R-R_I\\
\end{array}
\label{eq:cpara}
\end{equation}

\noindent where $R_X \equiv A_X / E(B-V)$, and $A_X$ is the dust
absorption in a given band. 
We note that $c_{BV}=1$ by construction (see Eq.\ref{eq:col}),
so we have 3 independent equations.  Fig.~\ref{mycardelli} shows $A_{\lambda}-A_B$ as a
function of wavelength for $E(B-V)=0.1$. The data points are determined using the $c$
parameters. The dotted line is the Cardelli, Claython \& Mathis 
(CCM) law assuming $R_V=3.1$ and the light grey solid line is the
result of fitting the CCM law on our data points.  
The best fit is obtained for $R_V=1.01 \pm 0.25$. The
dashed line is the result of a similar procedure obtained by
\citep{2007A&A...466...11G} while developing the {\salt} templates.
A typical value of $E(B-V)=0.1$ was used in analogy with the work
of \citet{2007A&A...466...11G}.
We note that our measurements agree well with the results from {\salt}
in the $R$ and $I$-bands, but disagree in the $U$-band. We note also,
that the {\salt} curve is not easily modeled by simply assuming
a different value for $R_V$ in CCM.

For testing the effect of selecting a low extincted sample, we have
introduced in the analysis 11 high extincted SNe.
We assumed the same colour curves obtained on the low extincted sample,
and we investigated the reddening law on the larger sample.
The $c$ parameters obtained in this way are
slightly different, and lead to a larger value of $R_V$ when fitted by the CCM law,
equal to $R_V=1.75 \pm 0.27$. Given that the two samples are not 
independent from each other, since 69 SNe are present in both samples,
the different values found are strikingly incompatible.

\section{Monte Carlo simulations}
\label{sec:MC}

Due to the potential correlation of the fitted parameters $a$, $b$ and
$c$, and the possible correlation in the intrinsic dispersion, we have chosen
to use Monte Carlo simulations to estimate the uncertainties in the
above parameters, as well as on $R_V$.  

The data suggests that the intrinsic colour dispersion is correlated 
between points at different epochs, e.g. if a SN is bluer than average 
at a given epoch, it is likely to remain bluer.\footnote{We note that,
the colour excess $E(B-V)$ determined from the $B-V$ colour, is used for
the fit of Eq.~\ref{eq:col} to all the other colours. Thus, a SN can
still remain redder or bluer than average even after the reddening is
taken into account. The reddening law that minimizes the scatter
around all the colours is the result of the method used.}

However, the  correlation coefficient as a function of epoch 
is quite difficult to quantify, 
given the small sample and the sizable measurement uncertainties.
Moreover, the inhomogeneity of the data sample could partially
introduce such correlations, especially in the $U$-band, for which
a false correlation could originate by problems in the
$S$-corrections between filter systems (see also the discussion in the next section).
For these reasons, when running the Monte Carlo simulations, we considered
the uncertainty in the data for the two extreme cases of either 
full correlation or no correlation at all in the intrinsic dispersion (see 
Section~\ref{sec:intrinsic}). We found the case with no correlation 
gives the larger uncertainties on the fitted parameters, and we 
have conservatively chosen these as our estimate of the uncertainties.

We generated 100 synthetic data samples, with as many lightcurves as
in the real set, according to the following prescription:
\begin{itemize}

\item We generate data points normally distributed around the light 
curves, with a dispersion, $\sigma$, given by the measurement
uncertainties, assumed to be uncorrelated, plus an intrinsic dispersion  
fully correlated in time, i.e. all
points of the same SN for a given band where given the same intrinsic
dispersion at all epochs (case {\it A}) and fully uncorrelated (case
{\it B}). The intrinsic dispersion is considered
Gaussian, with a standard deviation of $0.05/\sqrt 2$ \footnote{The value of 
0.05 has been chosen a posteriori as 
average intrinsic dispersion in all colours at
all epochs, given the results shown in Table~\ref{intrdisp}. 
Using a smaller value, such as 0.03, closer to the lower limit values
in Table~\ref{intrdisp}, we find smaller uncertainties on the parameters.
In addition, we assumed the intrinsic dispersion in colours to be equally distributed
in the two bands}. The data points are generated at the same epochs as
the real data. 

\item We compute the colours and use the measured $E(B-V)$ to add reddening 
in each colour, given an assumed value of $R_V$.

\item We fit the colour curves for the simulated samples, and retrieve the 
value of $R_V$ using the fitted $c$ parameters.

\end{itemize}

The dispersion in the fitted parameters in all the simulated data 
sets is taken as uncertainty in the parameters, $a$, $b$ and $c$ 
fitted on the real data, as reported in Tables~\ref{table:abUB}
-~\ref{table:abRI}. 

We run this test for two values: $R_V^{(true)} = 1$ and 
$R_V^{(true)} = 1.75$. In both cases we were able to retrieved the true value, 
i.e. we found no bias in our method. The dispersion measured as r.m.s. on the 
distribution of the $R_V$ can be taken as an estimated upper limit to the
uncertainties on the determination of $R_V$. We found $\sigma=0.13$
(case {\it A}) and $\sigma=0.27$ (case {\it B}). 

\section{Discussion}
\label{sec:discussion}

An extended analysis on supernova colour-curves and their dependence
on the light curve shape parameter was presented in previous
sections. 
An empirical model was defined, described by Eq.~\ref{eq:col}, and it
  was succefully fitted to the data. The robustness of the analysis, 
  was thoroughly tested. The residuals from the colour curves were
analyzed to search for further dependence on the stretch factor, SN
redshift, or colour excess $E(B-V)$. No significant correlation was
found.  Furthermore, the colour curves we derive describe well the
subsample of very low extinction objects, $E(B-V)\lsim 0.05$,
suggesting that the analysis is robust and not particularly affected
by moderately reddened SNe.  

In general, the dispersion we measured in all colours at all epochs is
non-negligible, indicating that colours of Type Ia supernovae have
more scatter than can be accounted for by the measurement
uncertainties.  These findings do not support earlier results 
based on a smaller sample where a negligible intrinsic scatter
at late epochs was found in some colours \citep{2003A&A...404..901N}.
The same conclusion was reached by
\citet{2007ApJ...659..122J} when studying an overlapping sample of
nearby SNe~Ia. They found an intrinsic dispersion of
$\sigma_{B-V}=0.062$ at epoch +35, in agreement with the result
reported here.  The intrinsic dispersion for various epochs and
colours is reported in Table~\ref{intrdisp} and could potentially be
used to constrain and discriminate between different models of the
physics of Type Ia SNe.  We notice, in general, a larger dispersion in
the $U-B$ colour at all epochs than in all other colours. Due to the
few data available, it is very difficult to establish whether the larger
dispersion is a real characteristic of the $U$-band or a consequence
of systematic effects. The $U$-band photometry is affected generally by larger
measurement uncertainties, smaller instrumental sensitivity, and larger
extinction corrections  (both atmospheric extinction, and 
interstellar extinction in the Milky Way and in the host galaxy). 
The broad variability of the effective $U$-band bandpass
shapes at different telescopes could lead to 
larger dispersion if accurate $S$-corrections are not applied (see
e.g appendix in \citet{2007A&A...469..645S}). Even larger are 
perhaps the uncertainties that could be introduced by the colour term in the
determination of the zero point. This is expected to be of the order
of 0.1 mag in the $U$-band photometry \citep{2000AIPC..522...65S}.
Many of these systematic uncertainties are correlated at different
epochs, and thus, can easily simulate a correlated intrinsic
dispersion as the one observed in the data. This is certainly a strong
concern for $U$-band, but it could be a second order effect for 
other bands as well. A larger and more homogeneous data sample, such
as the one collected by the SDSS collaboration, would be a
major improvement for investigating these hypotheses.

\begin{figure*}[htb]
\centering \includegraphics[width=9cm]{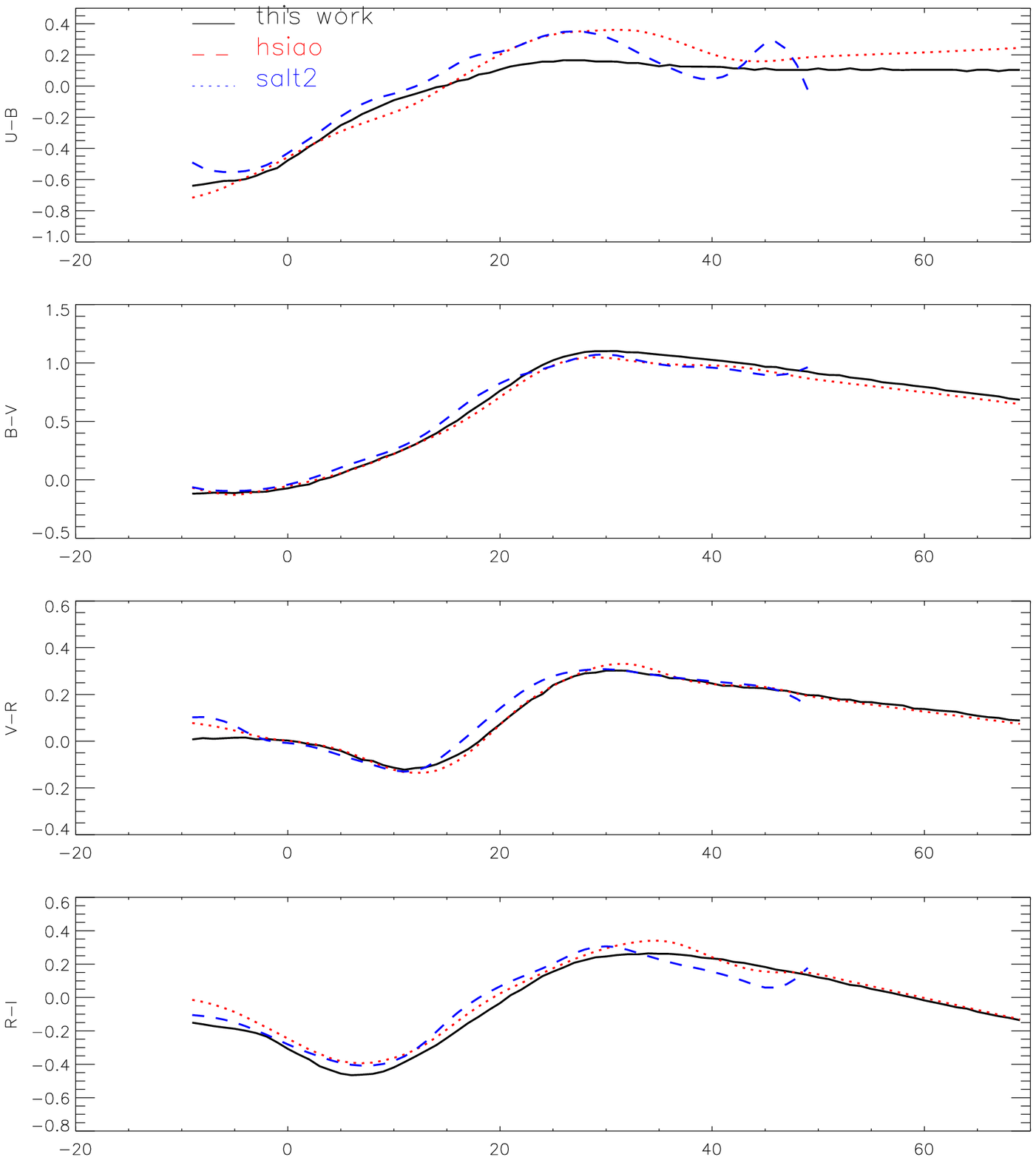}
\centering \includegraphics[width=9cm]{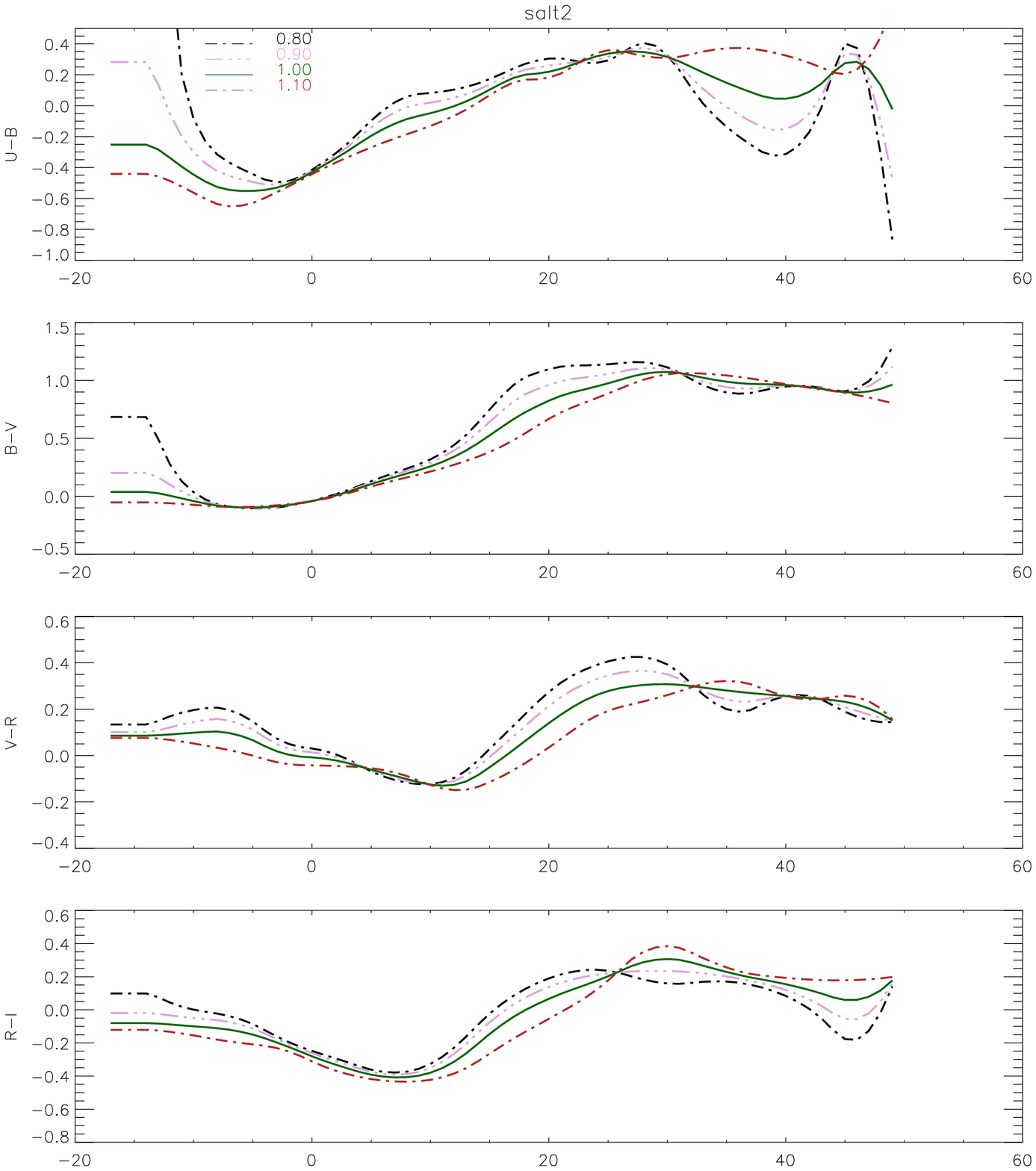}
\caption{Comparison of synthetic colour curves for all templates 
for $s=1$ supernova (left panel). Synthetic colour curve evolution for
different value of the stretch factor for the {\salt} templates (right panel).}
\label{comp_col}
\end{figure*}

We have built colour curves for {\hsiao} and {\salt} spectral
templates by computing synthetic photometry, and compared them with
our results.  The left-hand panel of Fig.~\ref{comp_col} shows a
comparison of the colour curves $U-B$, $B-V$, $V-R$ and $R-I$ for an
$s=1$ supernova as derived from the three spectral templates
considered. The right-hand panel shows the colour curves for different
stretch values for {\salt}.  We note that there is a general good
agreement in $B-V$ colour curves for a standard $s=1$ supernova, with
differences within the intrinsic colour dispersion. This is expected,
since the availability of both $B$ and $V$ band data is excellent.
The larger deviations between models are present in the $U-B$ colour
curve. We note in particular the strong dependence on stretch shown by
{\salt} around day +40. This is unexpected
and most probably an artifact of a poorly constrained
model due too sparse data in the $U$-band at
this late epochs. The same reasoning holds for the behaviour at early
epochs.  The {\salt} colour curves show overall more wiggles than the
colour curves derived in this work (see Fig.~\ref{models}). One
possible cause is the inhomogeneity of data quality. Some of the
supernovae in the sample have very small measurement errors 
and if no intrinsic color dispersion is considered, they 
weight very heavily in the overall fit, possibly biasing the model.
The comparison between spectral templates shown in
Fig.~\ref{spectra} leads to the same conclusions. The {\salt}
templates show a behaviour below 4000 {\AA} which is hard
to attribute to real spectral features, both for early and late time
(epoch=-10 and epoch=+20). These fluctuations are not present in the
{\hsiao} templates.  Another anomaly in the {\salt} templates is the
emission feature at $\sim 7600$ {\AA}, shown in many of the spectra
after maximum. Note, however, that these anomalies are within the
reported uncertainties in the {\salt} model. 

When comparing with other published results, one should keep
in mind the different aims of the various analyses. The {\salt}
templates, for instance, are a side product of a light curve
fitter. Thus, the colour curves derived by computing synthetic
photometry are not built on colour data, but are a consequence of
adjusting the spectra by the colours of the supernovae. In this
sense, they are subject to uncertainties introduced by the technique
used for warping the spectra.  Similarly the {\hsiao} templates are
built with the only aim of computing $K$-corrections, and not for
fitting lightcurves or exploring colour properties.  As mentioned in
section \ref{sec:template}, the average colour imposed to this
template is the one by \citet{2003ApJ...598..102K} which based the
$U-B$ colour on the measurements of five supernovae at maximum. 
Another important difference is the much larger number of 
parameters used for building {\salt} spectral templates, leading to
increased uncertainties 
compared to this work and the {\hsiao} templates.
When using few parameters, one assumes implicitly some ``resonable'' 
constrains on the models, such as smoothness   
and the number of knots for the spline interpolation.
The ability to relax these assumptions is limited by 
the quality and ammount of the avalaible data. 

In section~\ref{sec:extlaw}, we derive a reddening law from our sample
which is well described by the CCM law with a low value of
$R_V=1.01 \pm 0.25$ for SNe with $E(B-V)<0.25$ and $R_V = 1.75$ for
SNe with $E(B-V)<0.7$ mag. 
It should be noted that, we
have not used the CCM law in any part of the analysis, except for
correcting the data for galactic reddening. The spectral templates are
warped to match the average colour using a spline interpolation, and
not the CCM curve. We assumed no prior on any value of $R_V$ or on any
behaviour of the colour excess $E(B-V)$. Yet, we  find a good
match to the CCM law. The direct comparison with what found by
{\salt}, shows a very good agreement, except for the $U$-band
(see Fig.~\ref{mycardelli}), with differences of about 0.06 mag, for
$E(B-V)=0.1$. Given the intrinsic dispersion in the $U-B$, reported in
Table~\ref{intrdisp} this is not a large discrepancy. Moreover, besides 
all the concerns already discussed about possible problems in the 
$U$-band for the {\salt} templates, one should also
considered the additional uncertainties specific to 
$U$-band discussed above.

When adding the 11 highly reddened SNe in our sample, up to
$E(B-V) < 0.7$ mag\footnote {We note that a few SNe~Ia that 
show even higher reddening have recently been observed  
\citep{2006MNRAS.369.1880E,2008MNRAS.384..107E}. 
The values of $R_V$ determined for these individual SNe are in agreement with 
the larger value obtained in our work.}, the derived 
extinction properties change, yielding a larger value of the 
total-to-selective extinction parameter $R_V =1.75 \pm 0.27$. The two values are
statistically incompatible, given that they are determined using largely 
overlapping samples. Moreover, as pointed out in 
section~\ref{sec:MC}, the quoted uncertainty may be overestimated, since
determined assuming uncorrelated intrinsic dispersion between
SN epochs. A lower uncertainty is found for the fully correlated case, $\sigma_{R_V}\sim 0.13$, 
which would make the difference even more significant.

Using highly extincted SNe gives more leverage when studying
extinction law properties. As it is not possible to perform this
analysis only on the very small highly extincted SN sample available, the
determination of $R_V$ on the larger sample remains our most robust
global estimate of the total to selective extinction parameter.

Based on our findings we argue that it may not be surprising that
different analyses, based on overall minimization techniques of the
Hubble diagram, report different global fits of $R_V$, as its determination
depends much on the selection of the sample, and on the way the
uncertainties are estimated.  It is interesting to note that
\citet{1999ApJ...525..209T} found a very similar result when
minimizing the dispersion in the Hubble diagram: $R_V=0.876$ (inferred
from the measured $R_B=1.876$ when considering SNe with $E(B-V) <
0.20$ mag and $R_V=1.439$ (inferred from the measured $R_B=2.439$)
when this condition was relaxed.

As most samples used for fitting cosmological parameters, are selected 
based on low reddening, caution should be used when choosing 
the value of $R_V$ for reddening corrections. Even though the corrections
could be small, a systematic bias may be introduced, and could lead to substantially different
results, e.g. as shown by \citet{2007ApJ...664L..13C}.
 
The many studies reporting measurements of $R_V$ inconsistent with the
Milky-Way value ($R_V=3.1$) underline the need for a better
understanding of dust properties in other galaxies. In this work we
find that the reddening law for the low-extinction sample is
statistically incompatible with the value derived with the full
available sample. This may be taken as suggestive evidence for
additional processes involved in forming the effective reddening law
of SNe~Ia besides dimming by interstellar dust, e.g. scattered light 
echoes due to dust in the circumstellar environment
\citep{2005ApJ...635L..33W,2006MNRAS.369.1949P,2007arXiv0711.2570W}, 
or intrinsic properties of the SN explosion mimicking reddening by dust.
Unfortunatelly, only few high-quality colour measurements are currently 
available for highly reddened SNe~Ia. Additional observations of SNe~Ia 
with $E(B-V)>0.2$ should be carried out to further investigate this issue.

\begin{table}[htb]
\caption{$U-B$: time evolution of the $a$ and $b$ parameters.
$c_{U-B}=0.82 \pm 0.06$}
\begin{center}
\begin{tabular}{ccc}
\hline\hline
 Epoch & $b(t)$ & $a(t)$ \\
\hline
   -9  & -0.614 $\pm$  0.011  & -1.046 $\pm$  0.101  \\
   -6  & -0.594 $\pm$  0.011  & -1.002 $\pm$  0.090  \\
   -3  & -0.563 $\pm$  0.010  & -0.940 $\pm$  0.077  \\
    0  & -0.462 $\pm$  0.008  & -0.779 $\pm$  0.058  \\
    3  & -0.325 $\pm$  0.009  & -0.635 $\pm$  0.078  \\
    6  & -0.196 $\pm$  0.009  & -0.648 $\pm$  0.093  \\
    9  & -0.102 $\pm$  0.008  & -0.848 $\pm$  0.097  \\
   12  & -0.032 $\pm$  0.009  & -1.074 $\pm$  0.125  \\
   15  &  0.025 $\pm$  0.011  & -1.153 $\pm$  0.145  \\
   18  &  0.079 $\pm$  0.011  & -0.959 $\pm$  0.136  \\
   21  &  0.126 $\pm$  0.011  & -0.586 $\pm$  0.151  \\
   24  &  0.161 $\pm$  0.012  & -0.183 $\pm$  0.184  \\
   27  &  0.173 $\pm$  0.012  &  0.107 $\pm$  0.174  \\
   30  &  0.165 $\pm$  0.011  &  0.260 $\pm$  0.136  \\
   33  &  0.149 $\pm$  0.013  &  0.311 $\pm$  0.124  \\
   36  &  0.136 $\pm$  0.014  &  0.305 $\pm$  0.134  \\
   39  &  0.128 $\pm$  0.014  &  0.274 $\pm$  0.128  \\
   42  &  0.123 $\pm$  0.012  &  0.241 $\pm$  0.129  \\
   45  &  0.120 $\pm$  0.012  &  0.224 $\pm$  0.135  \\
   48  &  0.118 $\pm$  0.014  &  0.217 $\pm$  0.140  \\
   51  &  0.115 $\pm$  0.018  &  0.209 $\pm$  0.145  \\
   54  &  0.113 $\pm$  0.023  &  0.202 $\pm$  0.151  \\
   57  &  0.111 $\pm$  0.028  &  0.194 $\pm$  0.158  \\
   60  &  0.109 $\pm$  0.033  &  0.187 $\pm$  0.166  \\
   63  &  0.107 $\pm$  0.039  &  0.180 $\pm$  0.174  \\
\hline
\end{tabular}
\label{table:abUB}
\end{center}
\end{table}

\begin{table}[htb]
\caption{$B-V$: time evolution of the $a$ and $b$ parameters.
$c_{B-V}=1$ by construction.}
\begin{center}
\begin{tabular}{ccc}
\hline\hline
 Epoch & $b(t)$ & $a(t)$ \\
\hline

   -9  & -0.116 $\pm$  0.006  & -0.207 $\pm$  0.050  \\
   -6  & -0.104 $\pm$  0.005  & -0.187 $\pm$  0.045  \\
   -3  & -0.092 $\pm$  0.004  & -0.167 $\pm$  0.040  \\
    0  & -0.067 $\pm$  0.004  & -0.135 $\pm$  0.033  \\
    3  &  0.001 $\pm$  0.004  & -0.116 $\pm$  0.034  \\
    6  &  0.089 $\pm$  0.005  & -0.267 $\pm$  0.041  \\
    9  &  0.186 $\pm$  0.004  & -0.626 $\pm$  0.040  \\
   12  &  0.306 $\pm$  0.005  & -1.033 $\pm$  0.052  \\
   15  &  0.462 $\pm$  0.006  & -1.316 $\pm$  0.058  \\
   18  &  0.652 $\pm$  0.005  & -1.380 $\pm$  0.051  \\
   21  &  0.838 $\pm$  0.005  & -1.272 $\pm$  0.056  \\
   24  &  0.986 $\pm$  0.006  & -1.038 $\pm$  0.064  \\
   27  &  1.067 $\pm$  0.006  & -0.727 $\pm$  0.054  \\
   30  &  1.097 $\pm$  0.006  & -0.398 $\pm$  0.046  \\
   33  &  1.092 $\pm$  0.008  & -0.109 $\pm$  0.059  \\
   36  &  1.071 $\pm$  0.008  &  0.082 $\pm$  0.066  \\
   39  &  1.044 $\pm$  0.007  &  0.156 $\pm$  0.052  \\
   42  &  1.014 $\pm$  0.008  &  0.158 $\pm$  0.030  \\
   45  &  0.980 $\pm$  0.008  &  0.134 $\pm$  0.016  \\
   48  &  0.944 $\pm$  0.007  &  0.122 $\pm$  0.014  \\
   51  &  0.906 $\pm$  0.007  &  0.123 $\pm$  0.014  \\
   54  &  0.868 $\pm$  0.009  &  0.128 $\pm$  0.019  \\
   57  &  0.830 $\pm$  0.013  &  0.129 $\pm$  0.021  \\
   60  &  0.791 $\pm$  0.018  &  0.131 $\pm$  0.023  \\
   63  &  0.752 $\pm$  0.023  &  0.133 $\pm$  0.025  \\
\hline
\end{tabular}
\label{table:abBV}
\end{center}
\end{table}

\begin{table}[htb]
\caption{$V-R$: time evolution of the $a$ and $b$ parameters.
$c_{V-R}= 0.37 \pm  0.03$}
\begin{center}
\begin{tabular}{ccc}
\hline\hline
 Epoch & $b(t)$ & $a(t)$ \\
\hline
   -9  &  0.010 $\pm$  0.007  & -0.543 $\pm$  0.049  \\
   -6  &  0.008 $\pm$  0.006  & -0.513 $\pm$  0.044  \\
   -3  &  0.006 $\pm$  0.005  & -0.484 $\pm$  0.039  \\
    0  &  0.000 $\pm$  0.005  & -0.418 $\pm$  0.036  \\
    3  & -0.020 $\pm$  0.007  & -0.247 $\pm$  0.050  \\
    6  & -0.058 $\pm$  0.007  & -0.064 $\pm$  0.054  \\
    9  & -0.101 $\pm$  0.006  &  0.018 $\pm$  0.059  \\
   12  & -0.118 $\pm$  0.007  & -0.127 $\pm$  0.077  \\
   15  & -0.082 $\pm$  0.006  & -0.565 $\pm$  0.061  \\
   18  & -0.001 $\pm$  0.008  & -0.967 $\pm$  0.072  \\
   21  &  0.101 $\pm$  0.007  & -1.016 $\pm$  0.064  \\
   24  &  0.197 $\pm$  0.008  & -0.828 $\pm$  0.065  \\
   27  &  0.262 $\pm$  0.008  & -0.574 $\pm$  0.066  \\
   30  &  0.294 $\pm$  0.007  & -0.336 $\pm$  0.063  \\
   33  &  0.300 $\pm$  0.008  & -0.138 $\pm$  0.078  \\
   36  &  0.285 $\pm$  0.009  &  0.002 $\pm$  0.084  \\
   39  &  0.257 $\pm$  0.007  &  0.075 $\pm$  0.073  \\
   42  &  0.228 $\pm$  0.008  &  0.097 $\pm$  0.080  \\
   45  &  0.205 $\pm$  0.010  &  0.086 $\pm$  0.091  \\
   48  &  0.190 $\pm$  0.009  &  0.057 $\pm$  0.083  \\
   51  &  0.180 $\pm$  0.008  &  0.025 $\pm$  0.083  \\
   54  &  0.172 $\pm$  0.009  &  0.005 $\pm$  0.098  \\
   57  &  0.167 $\pm$  0.012  & -0.000 $\pm$  0.104  \\
   60  &  0.162 $\pm$  0.017  & -0.005 $\pm$  0.109  \\
   63  &  0.158 $\pm$  0.023  & -0.010 $\pm$  0.116  \\
\hline
\end{tabular}
\label{table:abVR}
\end{center}
\end{table}

\begin{table}[htb]
\caption{$R-I$: time evolution of the $a$ and $b$ parameters. 
$c_{R-I}= 0.40 \pm  0.03$ }
\begin{center}
\begin{tabular}{ccc}
\hline\hline
 Epoch & $b(t)$ & $a(t)$ \\
\hline
   -9  & -0.145 $\pm$  0.007  & -0.099 $\pm$  0.057  \\
   -6  & -0.167 $\pm$  0.006  & -0.104 $\pm$  0.051  \\
   -3  & -0.199 $\pm$  0.005  & -0.112 $\pm$  0.043  \\
    0  & -0.296 $\pm$  0.005  & -0.151 $\pm$  0.035  \\
    3  & -0.406 $\pm$  0.006  & -0.246 $\pm$  0.050  \\
    6  & -0.467 $\pm$  0.007  & -0.412 $\pm$  0.057  \\
    9  & -0.448 $\pm$  0.006  & -0.641 $\pm$  0.056  \\
   12  & -0.366 $\pm$  0.007  & -0.869 $\pm$  0.072  \\
   15  & -0.247 $\pm$  0.008  & -1.032 $\pm$  0.078  \\
   18  & -0.112 $\pm$  0.007  & -1.079 $\pm$  0.061  \\
   21  &  0.019 $\pm$  0.008  & -1.029 $\pm$  0.064  \\
   24  &  0.128 $\pm$  0.008  & -0.908 $\pm$  0.072  \\
   27  &  0.210 $\pm$  0.007  & -0.742 $\pm$  0.065  \\
   30  &  0.266 $\pm$  0.008  & -0.549 $\pm$  0.071  \\
   33  &  0.291 $\pm$  0.009  & -0.348 $\pm$  0.083  \\
   36  &  0.286 $\pm$  0.008  & -0.157 $\pm$  0.081  \\
   39  &  0.255 $\pm$  0.008  &  0.004 $\pm$  0.080  \\
   42  &  0.215 $\pm$  0.009  &  0.118 $\pm$  0.089  \\
   45  &  0.175 $\pm$  0.010  &  0.164 $\pm$  0.090  \\
   48  &  0.139 $\pm$  0.009  &  0.139 $\pm$  0.085  \\
   51  &  0.106 $\pm$  0.008  &  0.077 $\pm$  0.109  \\
   54  &  0.075 $\pm$  0.010  &  0.030 $\pm$  0.139  \\
   57  &  0.046 $\pm$  0.013  &  0.014 $\pm$  0.149  \\
   60  &  0.017 $\pm$  0.018  & -0.001 $\pm$  0.159  \\
   63  & -0.013 $\pm$  0.023  & -0.017 $\pm$  0.170  \\
\hline
\end{tabular}
\label{table:abRI}
\end{center}
\end{table}

\begin{acknowledgements}

The authors would like to thank
the G\"{o}ran Gustafsson Foundation 
and the Swedish Research Council for financial
support. We are very grateful to Rick Kessler for pointing out the
possibility of a bias in the determination of $R_V$. We would also like to
thank Julien Guy for helpful discussions about the {\salt} templates. 
Vallery Stanishev is acknowledged for stimulating discussions. 
We acknowledge the anonymous referee for her/his valuable comments, that helped 
improving the quality of the manuscript.

\end{acknowledgements}

\bibliographystyle{aa}
\bibliography{../../bibtex/bib}

\clearpage
\onecolumn

\longtab{1}{
\begin{longtable}{llccccc}
\caption{\label{listSNe} List of SNe used for the analysis.
References: (0), \citet{1994AJ....108.2233W}; (1), \citet{1996AJ....112.2398H}; 
(2), \citet{1998AJ....116.1009R}; (3), \citet{2006AJ....131..527J}; 
(4), \citet{2003AJ....125..166K}
(4a), \citet{2004AJ....127.1664K}; (4b), \citet{2004AJ....128.3034K}; 
(5), \citet{2003A&A...397..115V}; (6), \citet{2004MNRAS.355..178P}; 
(7), \citet{2007A&A...469..645S}; (8), \citet{2007MNRAS.tmp..161P}; }\\
\hline\hline
SN & Band &  $z$ & $s$ & $E(B-V)^{MW}$ & $E(B-V)^{HG}$ & Ref. \\
\hline
\endfirsthead
\caption{continued.}\\
\hline\hline
SN & Band &  $z$ & $s$ & $E(B-V)^{MW}$ & $E(B-V)^{HG}$ & Ref. \\
\hline
\endhead
\endfoot
    1989B &  U B V R I &  0.004 &  0.904 &  0.032  ( 0.003)   &  0.464  ( 0.021) &      (0) \\
    1990O &    B V R I &  0.030 &  1.072 &  0.093  ( 0.009)   &  0.061  ( 0.024) &      (1) \\
    1990T &    B V R I &  0.040 &  0.986 &  0.053  ( 0.005)   &  0.152  ( 0.021) &      (1) \\
    1990Y &    B V R I &  0.039 &  1.252 &  0.008  ( 0.001)   &  0.488  ( 0.026) &      (1) \\
   1991ag &    B V R I &  0.014 &  1.130 &  0.062  ( 0.006)   &  0.083  ( 0.014) &      (1) \\
    1991S &    B V R I &  0.056 &  0.945 &  0.026  ( 0.003)   &  0.024  ( 0.021) &      (1) \\
    1991U &    B V R I &  0.031 &  1.058 &  0.062  ( 0.006)   &  0.133  ( 0.022) &      (1) \\
    1992A &      B V R &  0.006 &  0.829 &  0.017  ( 0.002)   &  0.058  ( 0.006) &      (1) \\
   1992ae &        B V &  0.075 &  0.914 &  0.036  ( 0.004)   &  0.096  ( 0.024) &      (1) \\
   1992ag &      B V I &  0.026 &  0.820 &  0.097  ( 0.010)   &  0.267  ( 0.017) &      (1) \\
   1992al &    B V R I &  0.014 &  0.922 &  0.034  ( 0.003)   & -0.025  ( 0.011) &      (1) \\
   1992au &      B V I &  0.061 &  0.769 &  0.017  ( 0.002)   &  0.130  ( 0.034) &      (1) \\
   1992bc &    B V R I &  0.020 &  1.089 &  0.022  ( 0.002)   & -0.034  ( 0.009) &      (1) \\
   1992bg &      B V I &  0.036 &  0.996 &  0.185  ( 0.018)   &  0.051  ( 0.016) &      (1) \\
   1992bh &      B V I &  0.045 &  1.037 &  0.022  ( 0.002)   &  0.126  ( 0.015) &      (1) \\
   1992bl &      B V I &  0.043 &  0.771 &  0.011  ( 0.001)   &  0.015  ( 0.016) &      (1) \\
   1992bp &      B V I &  0.079 &  0.853 &  0.069  ( 0.007)   & -0.035  ( 0.017) &      (1) \\
   1992bs &        B V &  0.063 &  0.958 &  0.012  ( 0.001)   &  0.043  ( 0.015) &      (1) \\
    1992J &      B V I &  0.046 &  0.896 &  0.057  ( 0.006)   &  0.254  ( 0.028) &      (1) \\
    1992P &      B V I &  0.026 &  1.084 &  0.021  ( 0.002)   &  0.036  ( 0.017) &      (1) \\
   1993ac &    B V R I &  0.049 &  0.767 &  0.163  ( 0.016)   &  0.090  ( 0.040) &      (2) \\
   1993ae &    B V R I &  0.019 &  0.932 &  0.039  ( 0.004)   &  0.071  ( 0.013) &      (2) \\
   1993ag &      B V I &  0.050 &  0.909 &  0.112  ( 0.011)   &  0.132  ( 0.019) &      (1) \\
    1993B &      B V I &  0.071 &  0.823 &  0.079  ( 0.008)   &  0.078  ( 0.019) &      (1) \\
    1993H &    B V R I &  0.027 &  0.781 &  0.060  ( 0.006)   &  0.172  ( 0.014) &      (1) \\
    1993L &    B V R I &  0.005 &  1.159 &  0.014  ( 0.005)   &  0.457  ( 0.030) &      (1) \\
    1993O &      B V I &  0.052 &  0.898 &  0.053  ( 0.002)   & -0.012  ( 0.013) &      (1) \\
   1994ae &    B V R I &  0.004 &  1.006 &  0.031  ( 0.003)   &  0.102  ( 0.010) &      (2) \\
    1994D &    B V R I &  0.003 &  0.814 &  0.022  ( 0.002)   & -0.063  ( 0.000) &      (2) \\
    1994M &    B V R I &  0.023 &  0.800 &  0.024  ( 0.002)   &  0.093  ( 0.015) &      (2) \\
    1994Q &    B V R I &  0.029 &  1.131 &  0.017  ( 0.002)   &  0.155  ( 0.023) &      (2) \\
    1994S &    B V R I &  0.015 &  1.023 &  0.021  ( 0.002)   &  0.027  ( 0.019) &      (2) \\
    1994T &    B V R I &  0.035 &  0.928 &  0.029  ( 0.003)   &  0.061  ( 0.021) &      (2) \\
   1995ac &    B V R I &  0.050 &  1.077 &  0.042  ( 0.004)   &  0.057  ( 0.010) &      (2) \\
   1995ak &    B V R I &  0.023 &  0.831 &  0.043  ( 0.004)   &  0.109  ( 0.021) &      (2) \\
   1995al &    B V R I &  0.005 &  1.061 &  0.014  ( 0.001)   &  0.169  ( 0.013) &      (2) \\
    1995d &    B V R I &  0.007 &  1.069 &  0.058  ( 0.006)   &  0.042  ( 0.011) &      (2) \\
   1996ab &        B V &  0.124 &  1.006 &  0.032  ( 0.002)   & -0.024  ( 0.023) &      (3) \\
   1996bk &    B V R I &  0.007 &  0.763 &  0.018  ( 0.002)   &  0.380  ( 0.017) &      (3) \\
   1996bl &    B V R I &  0.036 &  0.982 &  0.105  ( 0.011)   &  0.077  ( 0.013) &      (3) \\
   1996bo &    B V R I &  0.017 &  0.949 &  0.078  ( 0.008)   &  0.323  ( 0.007) &      (3) \\
    1996C &    B V R I &  0.030 &  1.112 &  0.014  ( 0.001)   &  0.120  ( 0.016) &      (3) \\
    1996X &  U B V R I &  0.007 &  0.890 &  0.069  ( 0.007)   &  0.011  ( 0.006) &      (3) \\
    1996Z &      B V R &  0.008 &  0.819 &  0.063  ( 0.006)   &  0.425  ( 0.019) &      (3) \\
   1997dg &  U B V R I &  0.030 &  0.826 &  0.078  ( 0.002)   &  0.018  ( 0.015) &      (3) \\
   1997do &  U B V R I &  0.010 &  0.937 &  0.063  ( 0.002)   &  0.073  ( 0.010) &      (3) \\
    1997E &  U B V R I &  0.013 &  0.830 &  0.124  ( 0.002)   &  0.080  ( 0.006) &      (3) \\
    1997Y &  U B V R I &  0.017 &  0.875 &  0.017  ( 0.002)   &  0.031  ( 0.010) &      (3) \\
   1998ab &  U B V R I &  0.028 &  0.938 &  0.017  ( 0.002)   &  0.104  ( 0.008) &      (3) \\
   1998bu &  U B V R I &  0.003 &  0.953 &  0.025  ( 0.003)   &  0.344  ( 0.004) &      (3) \\
   1998dh &  U B V R I &  0.008 &  0.887 &  0.068  ( 0.002)   &  0.112  ( 0.009) &      (3) \\
   1998dx &  U B V R I &  0.054 &  0.818 &  0.041  ( 0.002)   & -0.048  ( 0.018) &      (3) \\
   1998ef &  U B V R I &  0.018 &  0.871 &  0.073  ( 0.002)   & -0.008  ( 0.009) &      (3) \\
   1998eg &  U B V R I &  0.024 &  0.992 &  0.123  ( 0.002)   &  0.060  ( 0.016) &      (3) \\
   1998es &  U B V R I &  0.010 &  1.074 &  0.032  ( 0.002)   &  0.143  ( 0.008) &      (3) \\
    1998V &  U B V R I &  0.017 &  0.930 &  0.196  ( 0.002)   &  0.033  ( 0.007) &      (3) \\
   1999aa &  U B V R I &  0.014 &  1.068 &  0.040  ( 0.002)   &  0.003  ( 0.005) &      (3) \\
   1999ac &  U B V R I &  0.010 &  1.111 &  0.046  ( 0.002)   &  0.091  ( 0.006) &      (3) \\
   1999cc &  U B V R I &  0.031 &  0.822 &  0.023  ( 0.002)   &  0.019  ( 0.017) &      (3) \\
   1999dk &  U B V R I &  0.015 &  1.042 &  0.054  ( 0.002)   &  0.155  ( 0.008) &      (3) \\
   1999dq &  U B V R I &  0.014 &  1.062 &  0.110  ( 0.002)   &  0.160  ( 0.005) &      (3) \\
   1999ef &  U B V R I &  0.039 &  1.033 &  0.087  ( 0.002)   & -0.007  ( 0.016) &      (3) \\
   1999ej &  U B V R I &  0.013 &  0.795 &  0.071  ( 0.002)   &  0.027  ( 0.019) &      (3) \\
   1999ek &  U B V R I &  0.018 &  0.914 &  0.561  ( 0.002)   &  0.179  ( 0.006) &      (3) \\
   1999gd &  U B V R I &  0.019 &  0.957 &  0.041  ( 0.002)   &  0.486  ( 0.013) &      (3) \\
   1999gp &  U B V R I &  0.027 &  1.212 &  0.056  ( 0.002)   &  0.126  ( 0.003) &      (3) \\
    2000B &  U B V R I &  0.020 &  0.970 &  0.068  ( 0.002)   &  0.229  ( 0.011) &      (3) \\
   2000ca &      U B V &  0.024 &  1.011 &  0.067  ( 0.002)   & -0.040  ( 0.007) &     (4a) \\
   2000ce &  U B V R I &  0.016 &  1.076 &  0.057  ( 0.002)   &  0.535  ( 0.009) &      (3) \\
   2000cf &  U B V R I &  0.036 &  0.931 &  0.032  ( 0.002)   & -0.001  ( 0.014) &      (3) \\
   2000fa &  U B V R I &  0.022 &  0.992 &  0.069  ( 0.002)   &  0.091  ( 0.008) &      (3) \\
   2001ba &        B V &  0.029 &  1.025 &  0.064  ( 0.002)   & -0.021  ( 0.009) &     (4a) \\
   2001bt &        B V &  0.015 &  0.875 &  0.065  ( 0.002)   &  0.232  ( 0.006) &     (4b) \\
   2001cz &      U B V &  0.015 &  1.007 &  0.092  ( 0.002)   &  0.146  ( 0.006) &     (4b) \\
   2001el &  U B V R I &  0.004 &  0.962 &  0.069  ( 0.002)   &  0.168  ( 0.003) &      (4) \\
    2001V &    B V R I &  0.015 &  1.120 &  0.020  ( 0.002)   &  0.128  ( 0.019) &      (4) \\
   2002bo &  U B V R I &  0.004 &  0.900 &  0.025  ( 0.002)   &  0.436  ( 0.006) &   (4b,5) \\
   2002er &  U B V R I &  0.009 &  0.896 &  0.160  ( 0.002)   &  0.195  ( 0.009) &      (6) \\
   2003du &  U B V R I &  0.006 &  0.991 &  0.010  ( 0.002)   & -0.069  ( 0.001) &      (7) \\
   2005cf &  U B V R I &  0.006 &  0.958 &  0.097  ( 0.002)   &  0.028  ( 0.002) &      (8) \\
\hline
\end{longtable}}

\end{document}